\documentclass[sn-nature]{sn-jnl}

\usepackage{graphicx}%
\usepackage{multirow}%
\usepackage{eufrak}
\usepackage[title]{appendix}%
\usepackage{xcolor}%
\usepackage{textcomp}%
\usepackage{manyfoot}%
\usepackage{booktabs}%
\usepackage{algorithm}%
\usepackage{algorithmicx}%
\usepackage{algpseudocode}%
\usepackage{listings}%
\usepackage{natbib}%
\usepackage{xfrac,bigints}
\usepackage{lmodern}
\usepackage{bm}
\usepackage{etoolbox}
\usepackage{hyperref}
\makeatletter
\patchcmd{\@maketitle}{\artauthors}{{\artauthors}}{}{}
\makeatother

\usepackage{xcolor}
\newcommand{\rev}[1]{{\color{black} {#1}}}

\begin{document}


\title[Article Title]{Active flow control for drag reduction through multi-agent reinforcement learning on a turbulent cylinder at $Re_D=3900$}

\author*[1]{\fnm{Pol} \sur{Suárez}}\email{polsm@kth.se}
\author[1]{\fnm{Francisco} \sur{Alcántara-Ávila}}
\author[2]{\fnm{Arnau} \sur{Miró}}
\author[3]{\fnm{Jean} \sur{Rabault}}
\author[4]{\fnm{Bernat} \sur{Font}}
\author[2]{\fnm{Oriol} \sur{Lehmkuhl}}
\author*[1]{\fnm{Ricardo} \sur{Vinuesa}} \email{rvinuesa@mech.kth.se}

\affil[1]{\orgdiv{FLOW, Engineering Mechanics}, \orgname{KTH Royal Institute of Technology}, \orgaddress{\city{Stockholm}, \postcode{SE-100 44}, \country{Sweden}}}

\affil[2]{\orgname{Barcelona Supercomputing Center (BSC-CNS)}, \orgaddress{\city{Barcelona}, \postcode{08034}, \country{Spain}}}

\affil[3]{\orgdiv{Independent Researcher}, \orgaddress{\city{Oslo}, \country{Norway}}}

\affil[4]{\orgdiv{Faculty of Mechanical Engineering}, \orgname{Technische Universiteit Delft}, \orgaddress{\city{ Delft}, \postcode{2600 AA}, \country{The Netherlands}}}

\abstract{This study presents novel \rev{drag reduction active-flow-control (AFC) strategies} for a three-dimensional cylinder immersed in a flow at a Reynolds number based on freestream velocity and cylinder diameter of \(Re_D=3900\). The cylinder in this subcritical flow regime has been extensively studied in the literature and is considered a classic case of turbulent flow arising from a bluff body. The strategies presented are explored through the use of deep reinforcement learning. The cylinder is equipped with 10 independent zero-net-mass-flux jet pairs, distributed on the top and bottom surfaces, which define the AFC setup. The method is based on the coupling between a computational-fluid-dynamics solver and a multi-agent reinforcement-learning (MARL) framework using the proximal-policy-optimization algorithm. \rev{This work introduces a multi-stage training approach to expand the exploration space and enhance drag reduction stabilization. By accelerating training through the exploitation of local invariants with MARL, a drag reduction of approximately \( 9\% \) is achieved. The cooperative closed-loop strategy developed by the agents is sophisticated, as it utilizes a wide bandwidth of mass-flow-rate frequencies, which classical control methods are unable to match. Notably, the mass cost efficiency is demonstrated to be two orders of magnitude lower than that of classical control methods reported in the literature. These developments represent a significant advancement in active flow control in turbulent regimes, critical for industrial applications.}
 }


\maketitle

\section{Introduction}\label{intro}

Active-flow-control (AFC) devices are essential tools across diverse industries, aiming to optimize fluid-flow processes, enhance performance, and improve overall efficiency~\citep{choi_review_2008}. Currently, the aeronautical sector \rev{needs} more robust and sophisticated systems to develop better control strategies. In this scenario, innovative solutions are required to address the pressing environmental concerns linked to fossil-fuel dependence. Discovering and understanding physical mechanisms to reduce air resistance \rev{are} crucial for the sustainable development of the transport industry. Passive-flow-control (PFC) solutions, while simpler and easier to integrate, typically lack the adaptability and performance capabilities of AFC methods. However, in critical sectors like aerospace, automotive, energy, and maritime, AFC devices emerge as pivotal tools, effectively managing airflow around surfaces, minimizing drag, boosting lift, and controlling separation. For instance, some passive systems are protuberances or fixed flaps such as vortex generators or winglets. On the other hand, active devices like slats and flaps placed along airplane wings enhance maneuverability and efficiency. \rev{Dynamically optimizing} all these devices is challenging due to the complex interactions between pressure and viscous effects across multiple flight conditions. To design and converge \rev{on} possible solutions, substantial experience and computational resources are required.

Recent advancements in flow control have been complemented by the integration of machine-learning (ML) techniques, offering significant promise to the aeronautics sector. This includes the exploration of fundamental issues in fluid mechanics~\citep{vinuesa_transformative_2023} and the development of novel approaches for both active and passive flow control (AFC and PFC)~\citep{Le_Clainche_aircraft_2023}. Deep reinforcement learning (DRL), particularly, has emerged as a rapidly expanding field within ML, capturing substantial interest. Following its success in domains like board games~\citep{alphago_2016} and robotics~\citep{ibarz_drl_robotics_2021}, DRL demonstrates effectiveness in systems where a controller interacts with an environment to optimize a particular task; note that this is a characteristic highly relevant to many AFC scenarios. In such instances, DRL can dynamically interact with the flow, receiving feedback and refining actions iteratively over time. Designing AFC setups involves \rev{working with complex}, high-dimensional systems, requiring significant computational power to explore the vast parameter space and identify \rev{optimal global values}. DRL and neural networks streamline this process, facilitating the development of effective control strategies with a reduced computational burden.

The state-of-the-art on DRL for AFC applications is rapidly expanding, featuring studies on flow control for two-dimensional (2D) cylinders across a range of $Re_D$ (Reynolds number based on inflow velocity $U_{\infty}$ and cylinder diameter $D$) from 100 to 8000, resulting in drag reductions of 17\% and 33\%, respectively~\citep{tang_robust_2020, LiZhang_2022, ren_applying_2021, chatzimanolakis2023drag, bernd_re2000_2023}.  \rev{DRL has also been tested against Linear Genetic Programming Control (LGPC) in a cylinder at \(Re_D = 100\), highlighting DRL's robustness against variable initial conditions and sensor noise, while LGPC provided compact and interpretable control laws~\cite{castellanos_2022}. In addition, it has been compared to other global optimization techniques~\cite{Pino_Schena_Rabault_Mendez_2023}.
}Specific studies have also focused on the mitigation of vortex-induced vibrations, \textit{e.g.}~\cite{chen_deep_2023}. Additionally, research on the application of DRL has been conducted on aircraft wings~\citep{vinuesa_flow_2022}, fluid-structure interaction~\citep{chen_deep_2023}, and controlling highly turbulent flows, as demonstrated in Ref.~\cite{Font2024}, successfully reducing a turbulent separation bubble at a friction Reynolds number of $Re_\tau=750$. There are \rev{also studies} on flow control in turbulent channels~\citep{guastoni_deep_2023} and Rayleigh--Bénard convection~\citep{vignon_effective_2023}. \rev{Recent literature~\citep{Wang_Karniadakis_2023} suggests the possibility of transfer learning from 2D cylinders to three-dimensional (3D) domains and higher \(Re_D\). Recent research~\citep{suarez2023active} has contributed to advancing the state-of-the-art in the control of 3D cylinders. This advancement involves DRL training directly in 3D, considering Reynolds numbers $Re_D=100$ to $400$ that include the transition to three-dimensional wake instabilities.} The latter \rev{uses} an AFC configuration comprising numerous zero-net-mass-flow (ZNMF) actuators managed through a multi-agent reinforcement-learning (MARL) framework. 

\rev{Although it is expected that the differences in flow physics between 2D and 3D flows would lead to even better results when training on full 3D physics, the increased complexity and unique characteristics of 3D flows also introduce challenges and opportunities for control that are are not encountered in simpler 2D configurations.}

\rev{The present work builds on previous successful training in transitional regimes, advancing further to tackle the significant challenge of achieving a subcritical Reynolds number of \(Re_D = 3900\). This represents a more complex scenario, marking the first exploration of such conditions in MARL state-of-the-art, with more intricate structures to analyze and learn from.} This classic case has been extensively investigated~\citep{lehmkuhl2013low,norberg1994experimental,parnaudeau2008experimental, ma_karniadakis_2000,kravchenko2000numerical, franke_2002}, serving as a reference for benchmarking and facilitating the study of well-known physics. Such insights are very valuable for devising an appropriate closed-loop control mechanism within a MARL framework. Despite the wealth of documentation available, consisting of numerous simulations and experiments, there remains a degree of inconsistency when comparing the time-averaged statistics in the near-cylinder wake. This inconsistency primarily stems from the high sensitivity to minor disturbances and the unsteady behavior of vortex formation, which directly impacts the configuration of the near wake. The primary point of discussion revolves around determining the number of shedding cycles required to attain converged statistics. Recent studies demonstrate how the presence of low-frequency fluctuation mechanisms, along with the well-established vortex-shedding frequency and smaller Kelvin--Helmholtz instabilities, contribute to the gradual contraction and expansion of the recirculation region~\cite{lehmkuhl2013low}.

We first considered a control periodic in time and uniform in the spanwise direction as a \rev{controlled reference case}. We identified the optimal frequency of actuation around the vortex-shedding frequency $f_{{\rm vs}}$, and also adjusted the maximum amplitude. Although this strategy led to drag reduction for $Re_D$ between 100 to 400~\cite{suarez2023active}, at the present Reynolds number of $3900$ this approach actually increased the drag by $30$ to $50\%$. 

Kim \& Choi in 2005 \cite{choi_distributed_2005} studied flow-control strategies for the 3D cylinder at $Re_D=3900$, and reported successful drag reduction control by considering two types of control: \textit{in-phase} and \textit{out-of-phase}. In their control strategies they consider sinusoidal profiles in the spanwise direction of the cylinder, but fixed blowing and suction constant in time. The velocity profile consists of a constant normal velocity of $\phi_{\rm max} = 0.1U_\infty$ over a \rev{jet width of $10^\circ$}. They assess various possible configurations by analyzing the impact of the spanwise wavelength $\lambda_z$ of their control. The difference between both control types is that, while the \textit{out-of-phase} has opposed blowing and suction on the top and bottom at the same spanwise location, the \textit{in-phase} has the same amount of blowing or suction for both surfaces. For a wavelength of $\lambda_z/D=\pi$ (hence, $\lambda_z/D=L_z$), they reported \rev{$25\%$ and $18\%$} drag reduction for the \textit{in-phase} and \textit{out-of-phase} cases, respectively. Since \textit{in-phase} does not comply with the ZNMF condition, in this study we will consider their \textit{out-of-phase} case as the \rev{controlled reference case and from now} will be denoted as KC05.

Developing flow-control strategies for fully turbulent 3D wakes around cylinders constitutes a significant challenge for DRL. As the wake becomes three-dimensional, the MARL setup \rev{must} effectively utilize spanwise characteristic structures to devise efficient control methods, which \rev{can} have profound implications for drag reduction. Note that DRL \rev{can} discover new strategies by \rev{maximizing} rewards $r_t$ for an agent interacting with the environment through actions $a_t$ and partial observations $s_t$. Through episodes of consecutive actions, neural-network weights \rev{are} updated, optimizing policies to maximize expected rewards. For recent advances in flow control using MARL, interested readers \rev{are} directed to Refs.~\cite{belus_exploiting_2019, brunton_closed-loop_2015, vignon_recent_2023}, where significant progress and insights \rev{have been} reported.

\section{Methodology}\label{sec:mets}

\begin{figure}
  \centering
  \includegraphics[width=\textwidth, trim={1cm 0 1cm 0}]{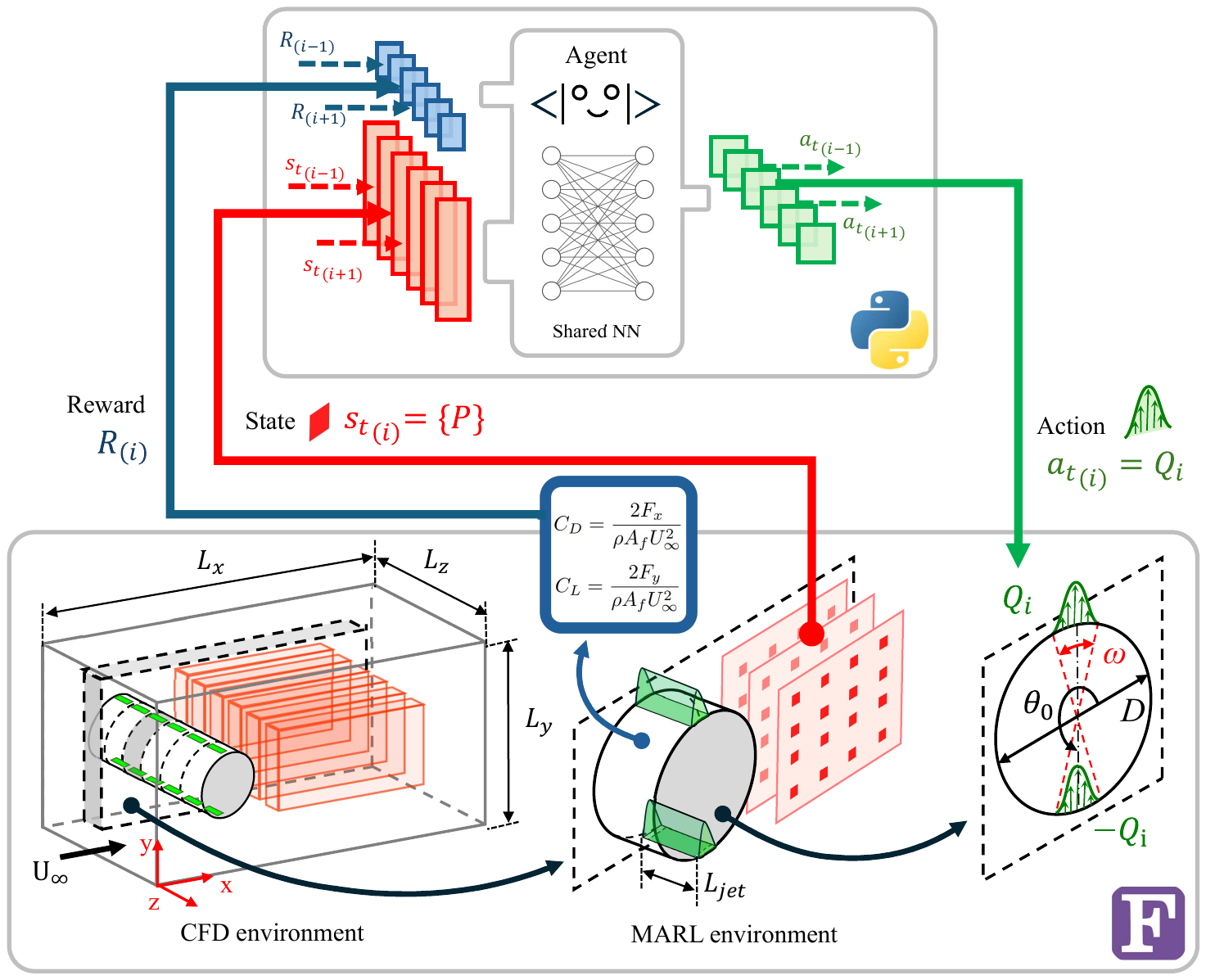}
  \caption{ Schematic representation that illustrates the multi-agent reinforcement-learning framework applied to a three-dimensional cylinder, showing communication channels between two main actors. In this case, the direction of the information is clockwise. At the top, we show the agent architecture featuring a shared neural network. At the bottom, the computational-fluid-dynamics (CFD) environment is depicted, with the cylinder diameter $D$ as the reference length. Moving rightward, emphasis is placed on the local MARL environment, also known as the pseudo environment. Note that $\omega$ denotes the jet angle width, while $\theta_0$ represents the angular location of each jet center. Additionally, the green shading illustrates the sinusoidal velocity profile, which remains uniform in the spanwise direction within a single jet length, $L_{\rm jet}$.}
  \label{fig:fig1}
\end{figure}

This study involves a 3D cylinder subjected to a constant inflow in the streamwise direction, with all lengths non-dimensionalized using the cylinder diameter $D$. The computational domain, depicted in Figure~\ref{fig:fig1}, has dimensions $L_x/D=40$, $L_y/D=25$, and $L_z/D=\pi$, with the cylinder centered at $(x/D,y/D)=(6.25,12.5)$. Here $x$, $y$ and $z$ denote the streamwise, vertical and spanwise directions, respectively. Note that the coordinate-system origin is located at the front face left-bottom corner. Periodic boundary conditions are used in the cylinder spanwise direction. As discussed in the references presented in Table \ref{tab:validate}, there is a consensus in the literature that a spanwise length of $\pi$ is sufficient to statistically capture all wavelengths of the relevant structures. At the inlet, a constant velocity $U_{\infty}$ is imposed with a Dirichlet condition. The surfaces of the cylinder include the no-slip and no-penetration conditions, while the top, bottom, and outflow surfaces of the domain box are set as zero-stress outlet. The cylinder incorporates two sets of $n_{{\rm jet}}=10$ synthetic jets positioned at the top and bottom surfaces ($\theta_0=90^{\circ}$ or $270^{\circ}$, respectively). \rev{Here, $L_{\rm jet}$ is defined as the spanwise length of the jets. Hence, the jet length is $L_{\rm jet}/D \simeq 0.314$, which is 21.5\% shorter than what was employed in previous studies at lower $Re_D$~\cite{suarez2023active}.}
 \rev{This will allow a more flexible strategy when controlling the spanwise structures in the wake, which are finer in the present higher-$Re_D$ case where mode B of $\lambda_z/D=1$ is dominant--instead of the mixed mode A ($\lambda_z/D=4$) and B experienced during transition regimes. The current setup provides at least two jets for each mode B structure, ensuring greater control authority.} As discussed in \S\ref{intro}, this setup resembles the one reported in Ref.~\cite{choi_distributed_2005}, with two key differences. \rev{First, it will not be a prescribed control, as the DRL framework enables dynamic adjustments within a closed-loop system. Second, the control will vary both in the spanwise direction and over time}. In the current study, the jet velocity profile in the direction normal to the surface is defined in terms of the angle $\theta$ and the desired mass-flow rate per unit length $Q$ as follows:

\begin{align}
\lVert {U_{\rm jet}}(Q,\theta)\lVert = Q\frac{\pi}{\rho D \omega }\cos \left [ \frac{\pi}{\omega}(\theta-\theta_0) \right ],
\label{eq:ujet}
\end{align}

\noindent where $Q = \dot{m}/L_{z}$, $|\theta-\theta_0|\in[-\omega/2, \omega/2]$, $\dot{m}$ is the mass flow rate and $\omega$ is the angular opening of the jet as shown in Figure~\ref{fig:fig1}. For every pseudo environment (also called MARL environment, as discussed later), we set opposite action values between the pair of top and bottom jets, {\it i.e.} $Q_{\rm{90^\circ}}=-Q_{\rm{270^\circ}}$, to guarantee an instantaneous global zero net mass flux, as discussed in Ref.~\cite{suarez2023active}.

The numerical simulations are carried out by means of the numerical solver Alya, which is described in detail in Ref.~\cite{vazquez_alya_2016}. The spatial discretization is based on the finite-element method (FEM) and the incompressible Navier--Stokes equations, which are formulated below:

\begin{align}
\partial_{t}\bm{u}+(\bm{u}\cdot \nabla)\bm{u}-\nabla \cdot(2\nu \bm{\epsilon})+\nabla p &= \bm{f},\\ 
\nabla \cdot \bm{u} &= 0, 
\end{align}
\label{eqn:NS1}

\noindent are integrated numerically. Here $\bm{u}$ is the velocity vector, $\nu$ is the fluid kinematic viscosity, $\bm{\epsilon}$ is the strain-rate tensor $\bm{\epsilon}=1/2(\nabla \bm{u} + (\nabla\bm{u})^\mathrm{T})$ and $\bm{f}$ represents external body forces. In Equation (\ref{eqn:NS1}), the convective term $(\bm{u}\cdot\nabla)\bm{u}$ \rev{is} formulated to conserve energy, momentum, and angular momentum, as described in Refs.~\cite{charnyi_conservation_2017, charnyi_efficient_2019}. Time discretization \rev{employs} a semi-implicit method where the convective term follows a second-order Runge--Kutta scheme, and a Crank--Nicholson scheme \rev{is} utilized for the diffusive term~\cite{crank_practical_nodate}. Alya \rev{determines} the suitable time step using an eigenvalue-based time-integration scheme~\cite{article_trias}. Subsequently, the numerical solution of these equations \rev{is} computed for each time step. Drag and lift forces ($F_x$ and $F_y$, respectively) \rev{are} computed through integration over the cylinder surface $s$:

\begin{equation} 
\bm{F}=\int (\bm{\varsigma} \cdot \bm{n}) \cdot \bm{e}_j \rm{d} s,
\label{eqn:force}
\end{equation}

\noindent where $\bm{\varsigma}$ is the Cauchy stress tensor, $\bm{n}$ is the unit vector normal to the surface, and $\bm{e}_j$ is a unit vector aligned with the direction of the main flow velocity for $F_x$ and the perpendicular cross-flow direction for $F_y$.

In order to define the uncontrolled case it is important to carefully study the convergence of the cylinder at $Re_D=3900$. In this study we use an \rev{unstructured mesh in the $xy$-plane, which is then extruded} in the $z$ direction. Following a convergence study, an interval of 300 convective time units, which are defined in terms of $U_{\infty}$ and $D$, $tU_{\infty}/D$, is considered to be sufficiently long to properly capture the pressure distribution around the cylinder, which in turn is associated with the computation of the aerodynamic forces, since the drag of a cylinder in these regimes mostly comes from the pressure component. \rev{Following the grid--independence study reported in~\ref{App1}, the simulations with the chosen mesh show reasonable agreement with the results reported in the literature for this case, as shown in Table~\ref{tab:validate}. Note that there is a discrepancy in $\overline{C_D}$ (defined later in Equation~(\ref{eq:aero_for}) in \S\ref{sec:MARL}) compared to the results of Lehmkuhl \textit{et al.}~\cite{lehmkuhl2013low}. This discrepancy may arise from the case being particularly sensitive in the near-cylinder region, where very low-frequency modulations associated with higher $\overline{C_D}$ have been also reported in the literature.}

\begin{table}[ht]
\caption{The statistical values for flow around a cylinder at $Re_D=3,900$ are presented, with comparisons made between the results of the present uncontrolled case and results reported in the existing literature.}
\label{tab:validate}
\begin{tabular}{cccccc}
\toprule
Reference & \textbf{$L_z/D$} & $St$ & $L_r/D^*$ & $\overline{C_D}$ & $-\overline{C_{pb}}^{**}$\\
\midrule
    Present uncontrolled case & $\pi$ & 0.22 & \rev{1.30} & 1.08 & \rev{0.95} \\
    Lehmkuhl \textit{et al.}~\cite{lehmkuhl2013low} & $\pi$ & 0.215 & 1.36 & 1.015 & 0.935 \\
    Parnaudeau \textit{et al.}~\cite{parnaudeau2008experimental} & 23 & 0.208 & 1.51 & ... & ... \\
    Norberg~\cite{norberg1998ldv} ($Re_D=3000$) & 67 & 0.22 & 1.66 & 0.98 & 0.88 \\
    Lourenco and Shih~\cite{Loureno1993CharacteristicsOT} & 21 & ... & ... & 0.98 & 0.9 \\
    Tremblay \textit{et al.}~\cite{tremblay2002} & $\pi$ & 0.22 & 1.3 & 1.03 & 0.93 \\
    Kravchenko \& Moin~\cite{kravchenko2000numerical} & $\pi$ & 0.21 & 1.35 & 1.04 & 0.94 \\
\botrule
\end{tabular}

\footnotetext[*]{The recirculation bubble length ($L_r/D$) is measured as the distance from the rear point of the cylinder to the position where the centerline velocity in the $x$-direction satisfies $\overline{u}=0$.}

\footnotetext[**]{The back pressure coefficient ($C_{pb}$) is defined as $C_{pb} = (\overline{p} - p_\infty) / \left(\frac{1}{2}\rho_\infty U_\infty^2\right)$.}

\end{table}

\subsection{Multi-agent reinforcement learning (MARL)}\label{sec:MARL}

We \rev{implement} a deep-reinforcement-learning (DRL) framework using the Tensorforce \rev{Python} library~\citep{schaarschmidt_tensorforce_2017}. DRL \rev{is} very well suited for unsteady flow-control problems. It \rev{provides} the possibility to dynamically interact with an environment in a closed-loop approach, \rev{being} able to set the actuation based on the varying flow state \rev{following a trained policy $\pi(a_t \lvert s_t)$--which describes the probability of selecting an action $a_t$ given the current state $s_t$}. \rev{We \rev{use} the proximal-policy-optimization (PPO) algorithm~\citep{schulman2017proximal}, which is a policy-gradient approach based on a surrogate loss function for policy updates to prevent drastic drops in performance. This algorithm \rev{exhibits} robustness, as it \rev{is} forgiving with hyperparameter initializations and \rev{can} perform adequately across a diverse range of RL tasks without extensive tuning. While PPO was selected for its stability and ease of use in this context, we acknowledge that other reinforcement learning algorithms, such as DDPG or SAC, could also be viable options.}

The neural-network architecture \rev{consists} of two dense hidden layers of 512 neurons each. The batch size $M$, \textit{i.e.}, the total number of streamed experiences that the PPO agent utilizes for each gradient-descent iteration, \rev{is} configured to 60, exceeding the values employed in previous 2D cylinder experiments \citep{varela_deep_2022} and previous 3D training scenarios \cite{suarez2023active}. The main constraint to set such a value \rev{lies} in having 10 actuators per environment, requiring 10 streamed experiences which are synchronized. \rev{Thus, we must operate with a total of $n_{\rm jets} \times n_{\text{envs}}$ sets of experiences, similar to what has been reported in Ref.~\cite{rabault_accelerating_2019}.} A streamed experience \rev{encompasses} a collection of states, actions, rewards, and the predicted state that the agent anticipates achieving, \rev{denoted as $\{s_t, a_t, R, s_t'\}$ for each pseudo environment}. Moreover, we encounter computational resource limitations. If the batch size, $M$, \rev{is} excessively large, a single training session \rev{might} be interrupted before any batch update occurs, resulting in the \rev{loss of already explored trajectories.}

In previous studies on 2D cylinders, the different training stages \rev{are} executed using a single-agent reinforcement learning (SARL) setup. However, considering the effectiveness of MARL in managing multiple actuators simultaneously, as demonstrated in \rev{recent literature \cite{suarez2023active, vignon_effective_2023, guastoni_deep_2023, Font2024}}, SARL \rev{is} not a feasible choice for the current 3D cylinder configuration with distributed input forcing and distributed output reward (referred to as the DIDO scheme). As opposed to SARL, \rev{the MARL framework mitigates the curse of dimensionality by exploiting invariances and focuses on training local pseudo environments with the option for collaboration among agents to achieve a global objective.} This approach \rev{makes} high-dimensional control manageable, as the agents \rev{are} trained in smaller domains to maximize local rewards. \rev{This makes the problem more scalable as long as the size of stacked local invariants is maintained.} All agents \rev{share} the same neural-network weights, significantly accelerating training. Each pseudo environment \rev{is} linked to a pair of jets that actuate independently. \rev{Observation states $s_{t,i_{\rm jet}}$} \rev{comprise} partial pressure values along the domain, focused on the wake and near-cylinder regions to exploit the spanwise pressure gradients when controlling. As detailed in Table \ref{tab:case_params}, these pressure values \rev{form} multiple slices in the $xy$ plane, evenly spaced in the spanwise direction \rev{by $\Delta z_{\rm slice}/D = \pi/30$}. Each set of three slices \rev{corresponds} to an individual pseudo environment. The total reward $R(t,i_{\text{jet}})$, as defined in Equation (\ref{eq:reward_eq}), \rev{comprises} \rev{the sum of local, $r_{\rm local}$, and global, $r_{\rm global}$, rewards corresponding to each jet $i_{\text{jet}}$}. The scalar $K_R$ \rev{adjusts} the values approximately within the range $[0,1]$, while $\beta$ \rev{balances} the local and global rewards; in this work, $\beta=0.8$. The rewards $r(t, i_{\rm jet})$, defined in Equation (\ref{eq:reward_cd}), \rev{depend} on aerodynamic force coefficients \rev{$C_D$ and $C_L$ ($\overline{C_{D_{\text{b}}}}$} \rev{represents} the averaged value for uncontrolled conditions). The user-defined parameter $\alpha$ \rev{serves} as a lift penalty, and in our study we \rev{set} $\alpha=0.6$. This parameter \rev{is} crucial for mitigating undesired asymmetric strategies that favor a reduction of the component parallel to the incident velocity (drag) over the perpendicular one (positive or negative lift), commonly known as the axis-switching phenomenon. Note that Table~\ref{tab:case_params} \rev{summarizes} the rest of MARL and computational-fluid-dynamics (CFD) parameters that \rev{define} the whole framework employed here.

\begin{gather}
    \label{eq:reward_eq} R(t,i_{\rm{jet}})=K_R \left [\beta r_{\rm{local}}(t,i_{\rm{jet}}) + (1-\beta)r_{\rm{global}}(t) \right ], \\
    \label{eq:reward_cd} r(t,i_{\rm{jet}})=\overline{C_{D_{\text{b}}}}-C_D(t,i_{\rm{jet}})-\alpha\vert C_L(t,i_{\rm{jet}})\vert,\\
    \label{eq:aero_for}\text{where} \quad C_D=\frac{2 F_x}{\rho A_f U_{\infty}^{2}}  \quad \text{and} \quad C_L=\frac{2 F_y}{\rho A_f U_{\infty}^{2}}.
\end{gather}

The aerodynamic forces described in Equation (\ref{eq:aero_for}) incorporate the frontal area $A_f = DL_{\rm jet}$, derived from the local pseudo-environment surfaces for \rev{$C_{D_{\rm{local}}}$}, and from the entire cylinder for $C_{D_{\rm{global}}}$.

\rev{The interactions between the agent and the physical environment are represented by actions \(a_t\), which influence the system over a time interval of \(T_a\) time units. We update each jet boundary condition using Equation (\ref{eq:ujet}) with its corresponding \(Q_{t,i}\). To transition smoothly in time between the actions at \(t\) and \(t+1\) \textit{i.e.}, \(Q_{t,i} \rightarrow Q_{t+1,i}\), we employ exponential space-time functions. This ensures a gradual shift in time, reducing the occurrence of sudden mass discontinuities that could disrupt the training process. These functions exhibit better performance than the linear slopes employed in Ref.~\cite{varela_deep_2022}. Regarding the spatial distribution of the \(\{Q_1, Q_2, \dots, Q_{n_{\rm jets}}\}_t\), Heaviside functions are used to activate or deactivate each jet depending on its location. Therefore, spatial smoothing has not been necessary for the present work. The DRL Python library requires normalization of the output actions \(a_t\) to the range \([-1, 1]\). To achieve this, the output actions are scaled by a factor \(Q_{\rm{max}}\), such that \(Q = a_t Q_{\rm{max}}\). Accordingly, \(Q_{\rm{max}} = 0.176\) was determined based on our experience with DRL for flow control in Refs.~\cite{suarez2023active,varela_deep_2022}.}

A vortex-shedding period is $T_{\rm{vs}}=1/St\approx4.7$ time units, \rev{based on our uncontrolled case and validated with the existing literature results}. Note that $St=f_{\rm{vs}} D/U_{\infty}$ is the Strouhal number, and $f_{\rm{vs}}$ denotes the vortex-shedding frequency. The episode duration is specifically set to span at least seven vortex-shedding periods ($T_{\rm{vs}}=1/f_{\rm{vs}}$). We choose $T_a\approx0.05T_{\rm{vs}}$, based on insights gained from previous studies \citep{rabault_artificial_2019}, \rev{\textit{i.e.} $T_a=0.25$ time units}. This interval allows sufficient time between actions to produce an effect on the flow. A shorter $T_a$ could introduce noise into the training process, complicating trajectory exploration and correlation \rev{within the $s_t$ gradients}. Conversely, an excessively long $T_a$ may compromise the capability of the agent to control shorter characteristic time scales. \rev{Thus, a total of 150 actuations \textit{i.e.} 37.5 time units per episode, seemed adequate for evaluating cumulative rewards, based on a preliminary estimation. 

At the beginning of the training, each episode starts from an uncontrolled, converged baseline state, with subsequent episodes beginning always from the last timestep of the baseline. After this first stage of the training period, the process is paused to evaluate the policy $\pi$ in exploitation mode, labeled DRL-10-s1 (stage 1).

Previewing the results discussed in \S\ref{sec_explotaition}, we observed that the episode duration for the DRL-10-s1 policy was more limiting than expected at this $Re_D$, restricting its ability to stabilize control performance. In short, the DRL-10-s1 case would lose and then recover drag control once it exceeded the predefined episode duration, indicating that it could not continue performing effective control beyond the episode duration. The reasoning behind this conclusion will be explained in detail later in \S\ref{sec_explotaition}.

In this work, we introduce a novel multi-stage training approach to address these limitations. The current training continues with a second stage, where a ratio $\delta$ of episodes begins from the last timestep of the previous episode, allowing for improved stabilization beyond the transition from uncontrolled to controlled state. This model, trained with the updated approach, is referred to as DRL-10-s2 (stage 2). Additionally, we compare our DRL-based control strategies with those developed using the Kim \& Choi \textit{out-of-phase} setup from 2005, here denoted as KC05.}

\begin{table}[ht]
\caption{Main parameters of the MARL architecture and the CFD setup used in the present work, for both DRL-10-s1 and DRL-10-s1 caseS.}
\label{tab:case_params}
\begin{tabular}{cc}
\toprule
\textbf{Parameter} & \textbf{Value/type} \\
\midrule
Number of grid points          &   $9.6 \times 10^6$    \\ 
$L_x/D$            &   40       \\   
$L_y/D$            &   25       \\   
$L_z/D$            &   $\pi$    \\
$L_{\rm{jet}}/D$   &  0.314     \\  
\rev{$s_t$ size}      & 183 (3 $xy$ slices of 61)     \\ 
\rev{$s_t$ variable}      & Pressure     \\
$Q_{\rm{max}}$     & 0.176                    \\
$K_R$            &       5             \\  
$\alpha$            &       0.6           \\
$\beta$            &       0.8           \\  
$T_a$                   &       0.25          \\
Actions/episode        &       150           \\   
CPUs/environment        &       1800          \\   
\rev{Parallel CFD environments $n_{\rm envs}$}  &       6             \\   
\rev{Actuators/CFD $n_{\rm jets}$}  &       10             \\
Batch size $M$  &       60             \\
Neurons (hidden layers)       &       512 (2)        \\    
Time-smoothing function         &       Exponential   \\ 
\botrule
\end{tabular}
\end{table}

\section{Results}\label{sec_results}

In this section, we present the successful training at $Re_D=3900$, which \rev{relies} on a MARL implementation. \rev{The proposed multi-stage training process, associated challenges, convergence assessment through reward evaluation, and its contributions, among other aspects, are described initially. Subsequently, both models trained, DRL-10-s1 and DRL-10-s2, are evaluated in exploitation mode, also known as deterministic mode.} The latter entails choosing actions without exploration; the agent solely applies the action associated with the highest probability of maximal reward given a particular state. Statistical results are presented alongside the uncontrolled case and the controlled reference case KC05. The purpose of conducting such a comparison \rev{is to clarify which novel drag-reduction physical mechanisms the agent explores.}

\subsection{Exploration}\label{sec_training}

Setting up a good training configuration in advance is crucial for achieving a consistent and efficient reward improvement. In fact, it is worth noting that to reach the configuration already shown in Table~\ref{tab:case_params}, there has been an iterative process of assessing different values of the main parameters. For instance, unsuccessful attempts were made with an observation state $s_t$ of $181\times3$ slices (543 pressure values in total), \rev{or taking into account more slices from neighbor actuators or even with $n_{\rm jets}=15$, hence shorter jets with $L_{\rm jet}/D \simeq 0.21$.} DRL requires methodical hyperparameter tuning in order to obtain \rev{the optimal setup} for the case under study.

\begin{figure}[ht]
    \centering
    \includegraphics[width=\textwidth, trim = 0.6in 1in 0in 0.75in, clip]{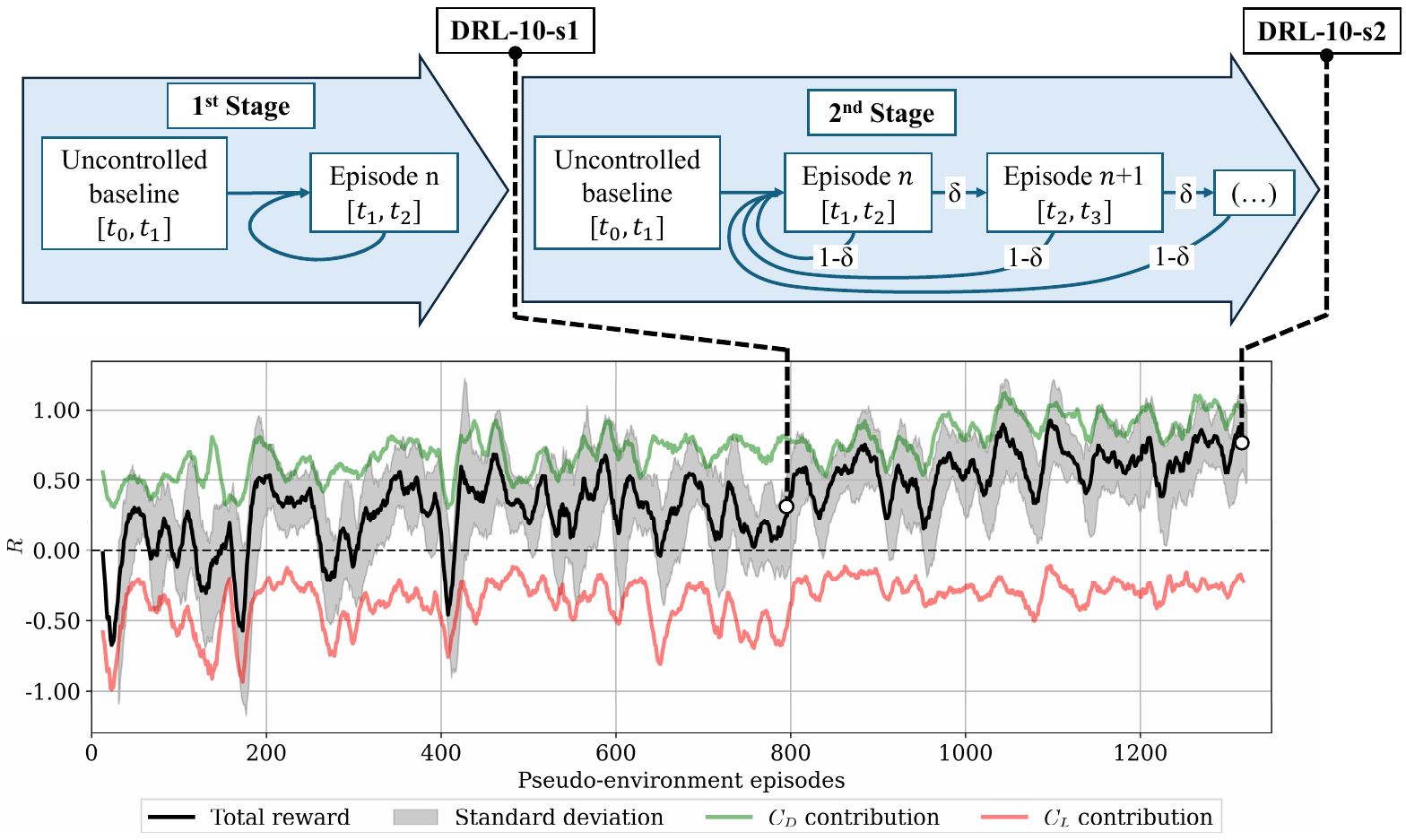}
    \caption{\rev{Evolution of all rewards collected at the end of pseudo-environment episodes, denoted as $R$, throughout the multi-stage exploration phase, along with its contributions from lift-bias and pure drag-reduction during the training session. The signals are smoothed using a moving average of 15 values window. Note that the top part describes the training of two stages.}}
    \label{fig:rewards}
\end{figure}

The exploration is evaluated continuously by monitoring the final and cumulative rewards in \rev{real wall-clock time}. Based on our experience, the most helpful metrics to track are the total reward $R$, its contribution due to lift-bias, $-\alpha\vert C_L(t,i_{\rm{jet}})\vert$, and the pure drag reduction, $\overline{C_{D_{\rm{b}}}}-C_D$. \rev{Additional data from Tensorboard logs, including policy and baseline losses, entropy, and episode returns, were analyzed to cross-validate our own metrics from CFD. However, tracking these metrics proved challenging due to multiple restarts from checkpoints and scattered data.} \rev{The reward evolution shown in Figure \ref{fig:rewards} highlights the difficulties when assessing whether a particular training is converged or not.} \rev{To explain and interpret this plot, which plays a crucial role in deciding whether to stop training early and thereby save computational resources, we observe two distinct trends corresponding to the two stages of training. This observation is noteworthy: during the final episodes of the initial stage, which lasts for 800 pseudo-environment episodes, exploration appeared stagnant, showing limited improvement—this posed a risk of overfitting, although such a risk is relatively low with PPO. However, with the workflow modifications introduced—outlined in the scheme at the top of the figure—the agent is encouraged to improve its ability to stabilize and achieve further improvements. As a result, the agent nearly doubled the total reward and reduced lift penalties over an additional 550 episodes.

At this point, we decided to stop the training process for two reasons: first, the reward had converged; second, the drag--reduction contribution approached our target value of \( R=1 \), which was predefined based on the KC05 results, indicating that the performance can be already considered satisfactory.

If we examine the details of the reward fluctuations, we observe that, although the reduction in $C_D$ converges to an acceptable reward value, the maximum, minimum, and standard deviation still contain significant levels of stochasticity, associated with the learning process and corresponding random exploration. To further assess the learning, intermediate exploitation of the model is also needed to monitor the drag reduction and if the control strategies are depicting any pattern. Based on our experience, this is very important when tackling unsteady and chaotic environments -- because such reward evolution can be misleading when deploying $\pi$. For instance, in a previous case with the model at 500 episodes (not reported here), the performance was similar to that after 800 episodes (DRL-10-s1). This suggests that training could have potentially been stopped earlier to allow for further adjustments. However, there is always a risk that the agent may not explore enough to reject poor trajectories. Rejecting poor trajectories is just as crucial as learning favorable ones, as it prepares the agent for any disturbances it may encounter.}

In terms of computational expense, \rev{training constitutes the dominant part. The presented training session required around $1350$ trajectories, akin to executing $135$ numerical simulations with 10 pseudo environments each.} All exploration sessions were conducted on the Dardel machine at PDC, the high-performance-computing (HPC) center at KTH. These sessions operate across 90 nodes concurrently, each 15 nodes executing a single numerical simulation consisting of 10 simultaneous pseudo environments, totaling 60 pseudo environments. Each node is equipped with two AMD EPYC™ Zen2 2.25 GHz 64-core processors and 512 GB of memory. With each batch of 6 CFD simulations optimally requiring approximately 10 hours in this specific \rev{setup, the process involves a minimum of two weeks of continuous operation. This is equivalent to using 11,520 CPU cores running for $\simeq 3$} million CPU hours in total. It should be noted that making an accurate estimation for such training sessions is difficult, considering synchronization times, the necessary restarts between episodes, and data movement in memory and on disk. After deciding to conclude the exploration phase, we proceed to evaluate the deployment of the model and its performance during exploitation.

\subsection{Exploitation}\label{sec_explotaition}

\begin{figure}[ht]
    \centering
    \includegraphics[width=\textwidth, trim = 1cm 0 1cm 0, clip]{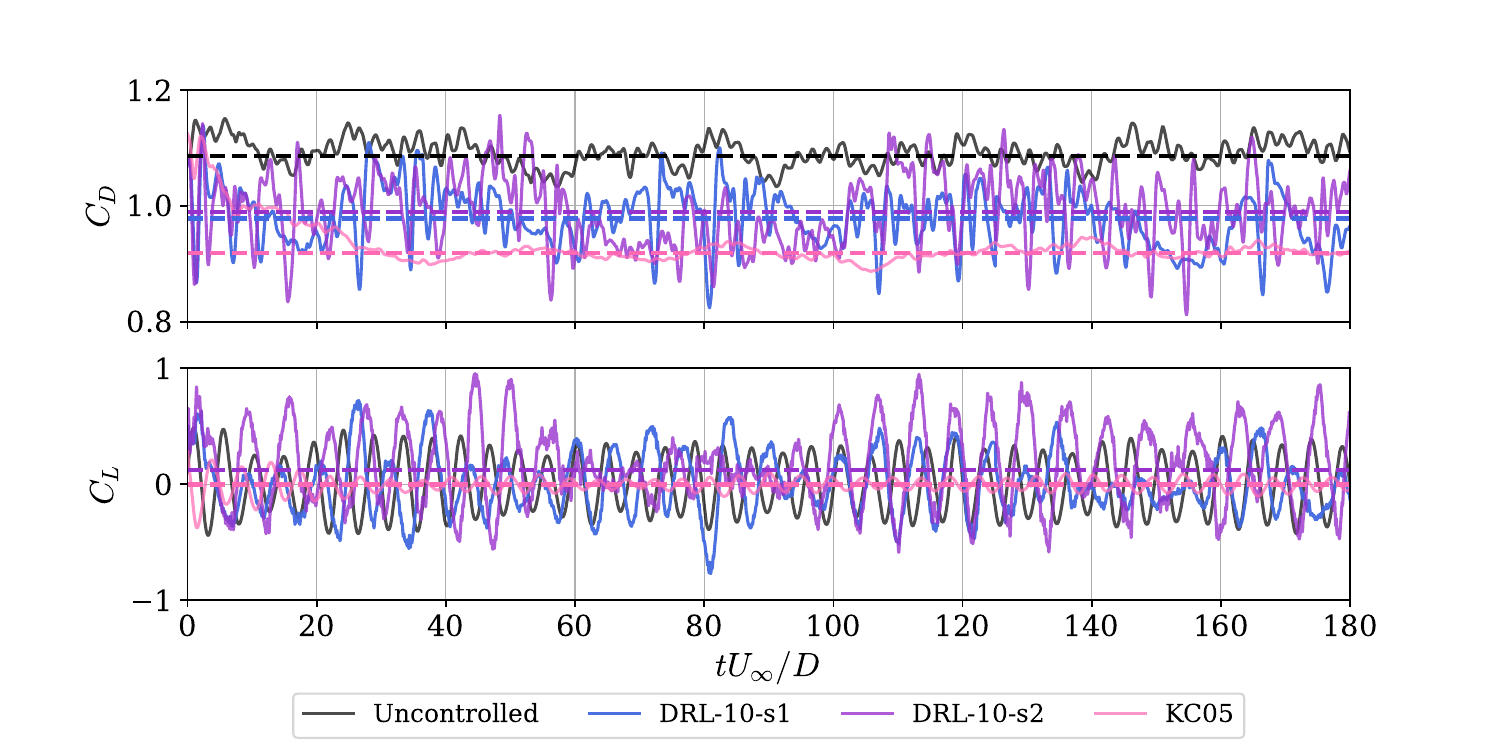}
    \caption{\rev{Evolution in time of the (top) drag coefficient $C_D$ and (bottom) lift coefficient $C_L$. Note that dashed horizontal lines are the averaged values and the transitional stages are included. When $tU_{\infty}/D=0$ the control starts for all cases, DRL-10-s1, DRL-10-s2 and KC05.}}
    \label{fig:cdandcl}
\end{figure}

When the agent operates without exploration, \rev{it always selects the best possible action from the policy $\pi$. In Figure \ref{fig:cdandcl}, we show how drag and lift coefficients evolves during exploitation for both models, DRL-10-s1 and DRL-10-s2}, alongside the control law from KC05 and the uncontrolled case. The effectiveness of the KC05 strategy is evident, but, as will be discussed later, it requires significantly higher actuation cost. \rev{Focusing on DRL-based strategies first, we observe a successful reduction of $C_D$. Regarding the results after the first stage, there is a reduction of $\overline{\Delta C_{D_{\rm s1}}}=-9.44\%$ with the DRL-10-s1 policy. Note that the drag exhibits some low-frequency oscillations with a period of approximately 40 time units---roughly the duration of the training episodes, spanning $t_1 \rightarrow t_2$, using nomenclature in Figure \ref{fig:rewards}. Within these intervals, the $C_D$ values transition quickly, in around 20 time units, to a brief but significant reduction of $\Delta C_D \approx -14\%$. However, the strategy is unable to sustain this reduction, and the drag forces revert to uncontrolled $C_D$ values and then back to the previous reduction again. This observation connects with the insights introduced earlier in \S\ref{sec:mets}. Our hypothesis is that the training episode duration is insufficient to fully capture the transition period and subsequently learn to stabilize this new flow state beyond these 40 time units.}

\rev{It is then that the idea of a second training phase emerges, in which the agent is forced to explore beyond the episode duration, aiming to reduce the oscillations that negatively impact average performance. After an additional 550 episodes of training, we run with the DRL-10-s2 strategy in exploitation mode, and the results partially confirm our expectations. We observe that the agent is now able to sustain lower $C_D$ values for a longer period, once again reaching $\Delta C_D = -14\%$ but now maintaining this reduction for over 40 time units--as seen between $tU_\infty/D=60$ to 100 in Figure \ref{fig:cdandcl}. However, it also appears to lose some ability to transition as quickly as before. Perhaps the $20\%$ rate of episodes starting from the uncontrolled case causes the agent to ``forget'' the transition process, trading it for improved stabilization of the reduction. Despite the improvement in low-frequency oscillations, the average reduction slightly worsens to $\overline{\Delta C_{D_{\rm s2}}}= -8.33\%$. Although DRL-based control occasionally surpasses KC05's drag reduction value of $\overline{\Delta C_{D_{\rm KC05}}} = -15\%$ (which is lower than the 18\% reported in the original study \cite{choi_distributed_2005}), the latter demonstrates consistency in control with fewer oscillations. This is also visible in the root-mean-squared values presented in Table \ref{tab:results}, where the DRL-based RMS values are up to three times higher than the uncontrolled case and five times higher than KC05. This resembles the behavior from the controlled cases studied in the transition regimes between $Re_D=100$ to $400$~\cite{suarez2023active}. On the other hand, a clear difference respect to the results in the transition regime is observed in the lift signals. Although the averaged values are also very close to zero, indicating no significant bias, the DRL-based strategies exhibit a minimal increase in fluctuations. In contrast, KC05 also yields good consistency with a low RMS. This suggest that the drag--reduction mechanisms explored here do not focus on minimizing these fluctuations in the pressure distribution.}

\rev{An important note to make at this point is that overfitting to a specific initial condition is discarded, as the DRL-based policies are tested with activation and deactivation at different timeframes. These tests show consistent performance with the results reported in Table \ref{tab:results}. However, these results are not reported in the present paper.}

Table~\ref{tab:results} also shows the rest of relevant physical quantities taking into account aerodynamic forces. All values are averaged over the last 150 time units in the converged stage, equivalent to $30$ shedding periods approximately. We observe how the DRL-based cases are able to influence $St$, while the KC05 case has no influence. The drag-reduction mechanism is very similar to those reported in Refs.~\cite{varela_deep_2022, suarez2023active}, at least statistically. Interestingly, the recirculation bubble is extended by \rev{41\% in DRL-10-s1, 49\% in DRL-10-s2 and by 67\% in KC05 relative to the case without control. The pressure valley at the most downstream point in the cylinder surface increases slightly, a fact that is directly related to the integral necessary for computing the forces exerted on the surfaces. However, although both DRL-based policies have the same value, it will be discussed below that the pressure distributions are different.}

The aspect in which DRL significantly outperforms classical AFC methods is in the cost associated with the control. Only considering the maximum values of mass-flow rate per unit length over time, it can be observed that it is an order of magnitude lower for both DRL-based cases compared to KC05. But if we also consider that KC05 maintains a constant value over time instead of fluctuating, we observe that $E_{c}^*/\Delta C_D$ ratios are two orders of magnitudes lower in the DRL case, compared with KC05. Note that $E_c^*$ is defined as: 
\begin{equation} \label{eq:Ec}
E_{c}^* = \frac{E_c}{E_\infty} = \frac{L_{{\rm jet}}}{(t_2 - t_1) Q_\infty L_z} \bigintss_{t_1}^{t_2} \sum_{i=1}^{n_{\rm{jets}}} \lvert Q_i(t) \lvert \ \rm{d} \textit{t},
\end{equation}

\noindent which represents the mass cost over time of any control system with $n_{\rm jets}$ compared to the corresponding mass intercepted by the cylinder $E_\infty$, where $Q_\infty=DU_\infty$. Note that low values of $E_c^*/\Delta C_D$ imply that control is much more efficient in terms of used mass per drag reduction $\Delta C_D$. 

Such good performance indicates that the DRL-based control is much more fine-grained than the fixed spanwise control by KC05, and is able to adapt to the instantaneous state of the wake, effectively exploiting the wake structures to achieve drag reduction. Note that in Table~\ref{tab:results} we also include the \rev{main frequency bandwidth} of the mass-flow control signal, denoted by $f_c$. It is interesting to note that this frequency differs from both the uncontrolled and the controlled Strouhal numbers.

\begin{table}[ht]
\caption{\rev{Summary of the statistical quantities for both DRL based controlled cases (DRL-10-s1 and DRL-10-s2) compared with the present uncontrolled case and KC05. All values are averaged over 150 time units after discarding the initial transients resulting after applying the control.}}
\label{tab:results}
\begin{tabular}{ccccc}
\toprule
     & \textbf{Uncontrolled} & \textbf{DRL-10-s1} & \textbf{DRL-10-s2} & \textbf{KC05} \\
\midrule
    $St$  & 0.22 & 0.177 & 0.172 & 0.22 \\
    $L_r/D$ Bubble length  & 1.30 & 1.80 & 1.94 & 2.18 \\
    $-\overline{C_{pb}}$  & 0.95 & 0.75 & 0.75 & 0.69 \\
    $Q_{\rm{max}}$  & ... & 0.053 & 0.023 & 0.11 \\
    $Q_{\rm{RMS}}$  & ... & 0.037 & 0.066 & ... \\
    $f_c$  & ... & 0.168-0.177 & 0.167-0.194 & ... \\
    $C_{D_{\rm{RMS}}}$ & 0.021 & 0.049 & 0.06 & 0.011 \\
    $C_{L_{\rm{RMS}}}$ & 0.238 & 0.25 & 0.34 & 0.048 \\ 
    $\overline{C_D}$    & 1.08 & 0.978 & 0.99 & 0.918 \\
    $\Delta \overline{C_D}$ [\%]  & ... & -9.44 & -8.33 & -15 \\
    $E_c^*/\Delta \overline{C_D}$ & ... & 0.0014 & 0.0015 & 0.22 \\
\botrule
\end{tabular}
\end{table}

\begin{figure}[ht]
    \centering
    \begin{minipage}[b]{\textwidth}
        \centering
        \includegraphics[width=\textwidth]{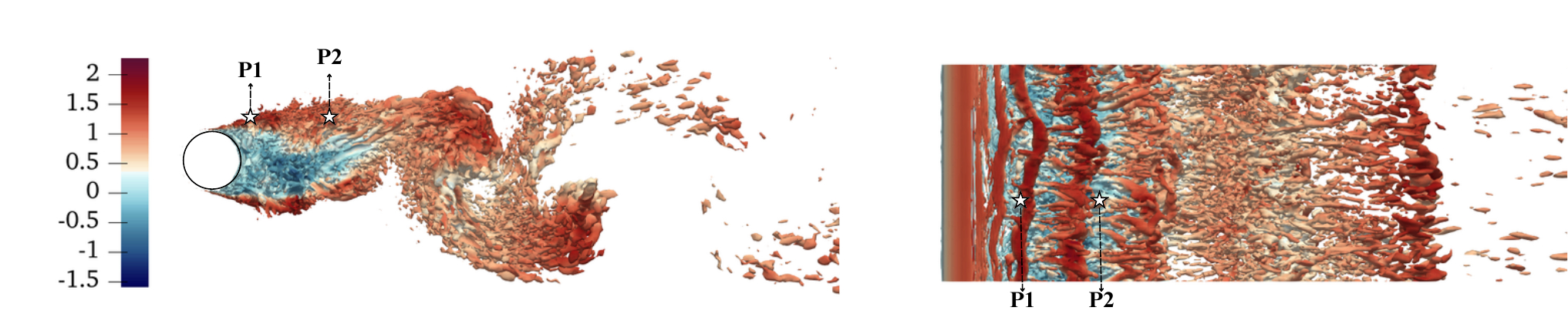}
        \begin{center}(a)\end{center}
        \label{fig:lamb_base}
    \end{minipage}
    \hfill
    \begin{minipage}[b]{\textwidth}
        \centering
        \includegraphics[width=\textwidth]{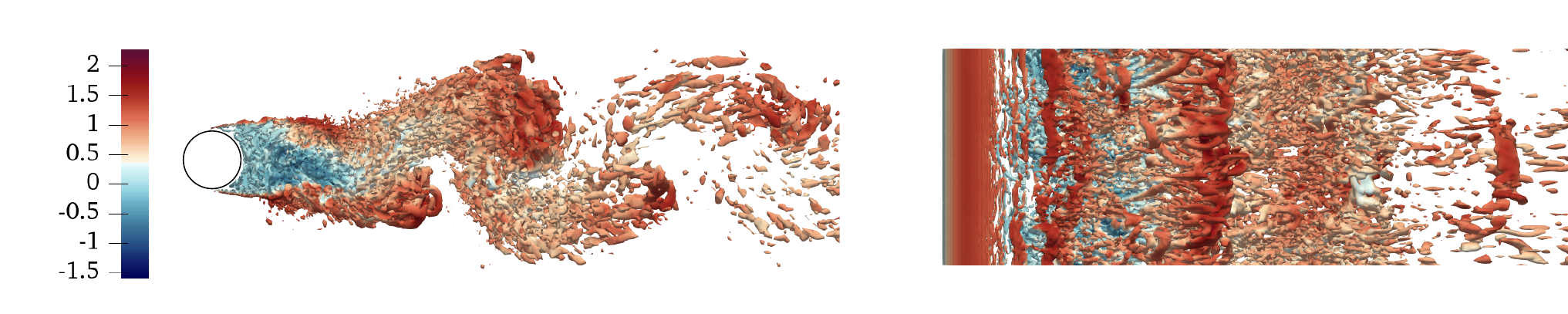}
        \begin{center}(b)\end{center}
        \label{fig:lamb_drl_s1}
    \end{minipage}
        \begin{minipage}[b]{\textwidth}
        \centering
        \includegraphics[width=\textwidth]{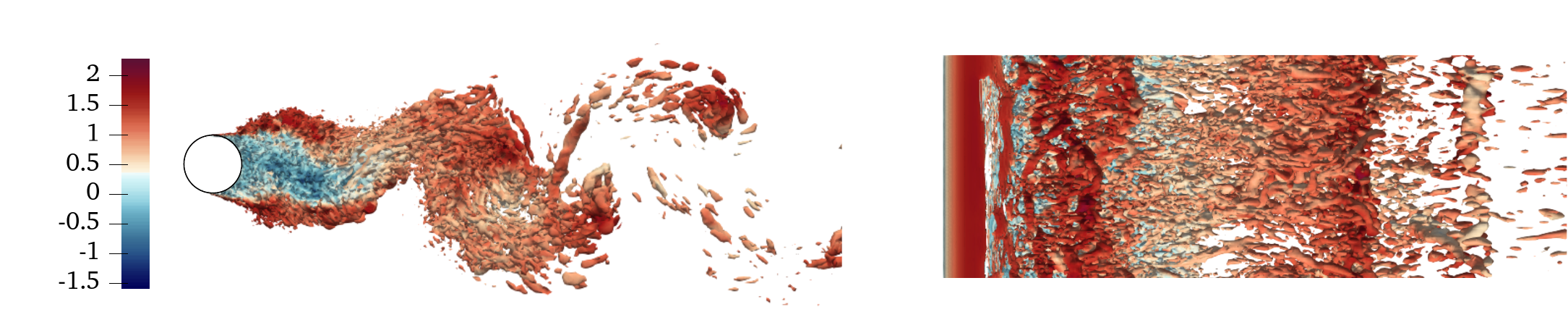}
        \begin{center}(c)\end{center}
        \label{fig:lamb_drl_s2}
    \end{minipage}
    \begin{minipage}[c]{\textwidth}
        \centering
        \includegraphics[width=\textwidth]{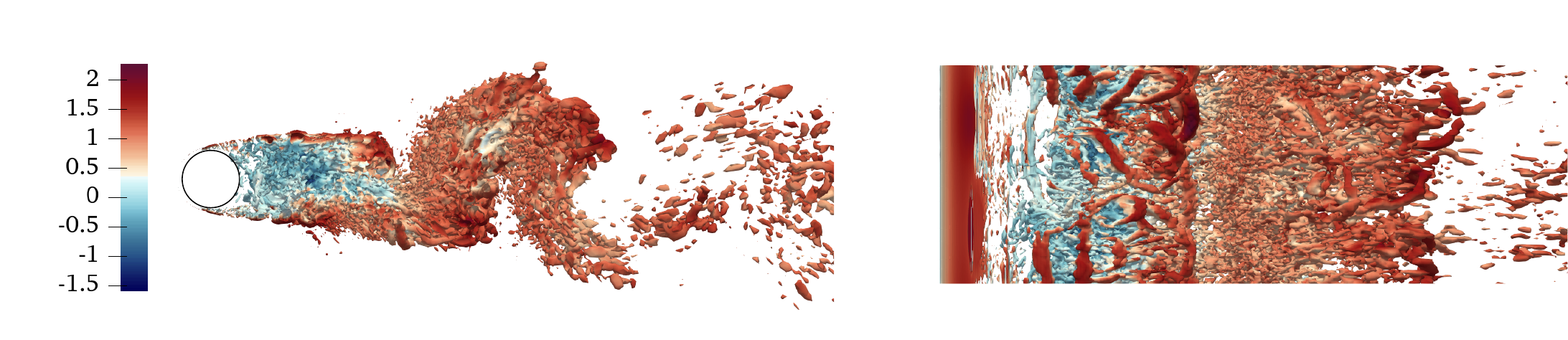}
        \begin{center}(d)\end{center}
        \label{fig:lamb_kim}
    \end{minipage}
    \caption{\rev{Instantaneous coherent structures identified with the $\lambda_2$ criterion~\cite{Jeong_Hussain_1995}, where the isosurfaces $\lambda_2 D^2/U_\infty^2=-5$ are coloured by the streamwise velocity. We show (a) uncontrolled, (b) DRL-10-s1, \rev{(c) DRL-10-s2}  and (d) KC05 cases, and all of them are chosen to show the flow in a statistically converged stage. Note that for (a) the locations of P1 and P2 that will be referred later in Figure \ref{fig:welch}(a) and \ref{fig:welch}(b) are shown.}}
    \label{fig:lamb}
\end{figure}

\rev{To study how the wake topology changes in the various scenarios, we present the vortical motions in the instantaneous flow in Figure~\ref{fig:lamb}. We note that similar patterns have been observed at different time sequences, although only one snapshot per case is shown here. The most remarkable aspect is the non-invasive nature of DRL-10-s1 and DRL-10-s2 compared to KC05, where a distinct peak of suction $z/D=\pi/4$ and another of blowing at $z/D=3\pi/4$ are clearly observed due to the wavelength covering the entire span of the cylinder. This is particularly evident in the Kelvin--Helmholtz instabilities, represented by the smaller tubular structures at the top, near the separation point, where the shear layer becomes unstable with high values of streamwise velocity--very visible in the $xz$-view in the right column. All controlled cases appear to extend these instabilities in the streamwise direction, helping to increase the recirculation bubble. 

However, there are some particularities: in the DRL-10-s1 case, a narrower shear layer is observed, but it is more aligned with the streamwise direction. In contrast, in stage 2, the shear layer is wider and forms a more open angle. In the case of KC05, the shear layer is notably more stable, maintaining the secondary tubular structures for a longer distance as part of the main vortex shedding process, thereby enlarging the recirculation bubble a bit more than the other cases. Additionally, all control strategies delay the emergence of the main counter-rotating vortices that initiate vortex shedding.}

\begin{figure}
    \centering
    \begin{minipage}[b]{\textwidth}
        \centering
        \includegraphics[width=\textwidth,trim = 0.0cm 0.5cm 0.1cm 0.5cm, clip]{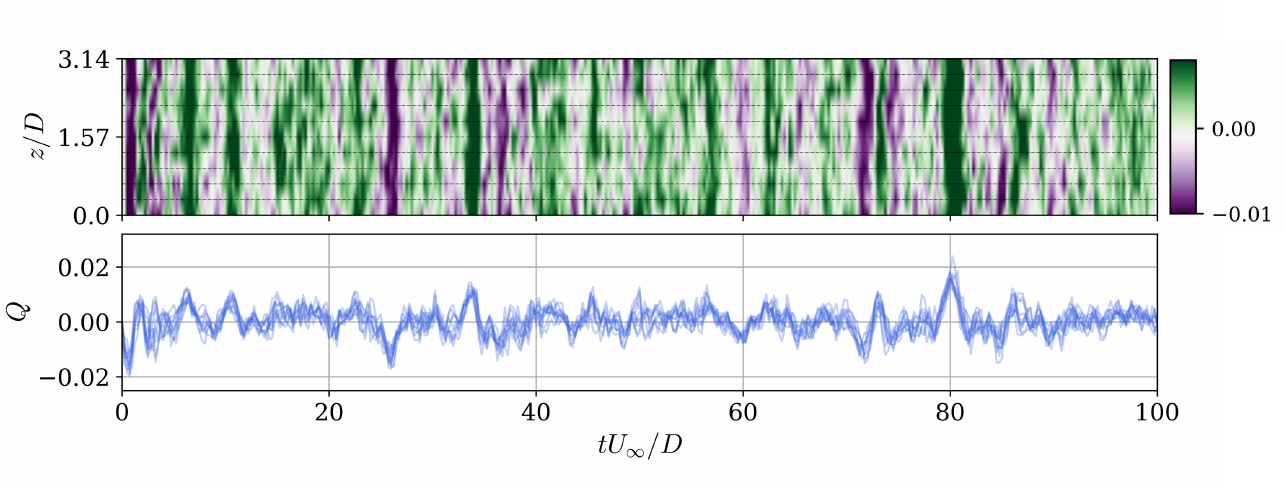}
        \begin{center}(a)\end{center}
    \end{minipage}
    \hfill
        \begin{minipage}[b]{\textwidth}
        \centering
        \includegraphics[width=\textwidth,trim = 0.0cm 0.2cm 0.1cm 0.2cm, clip]{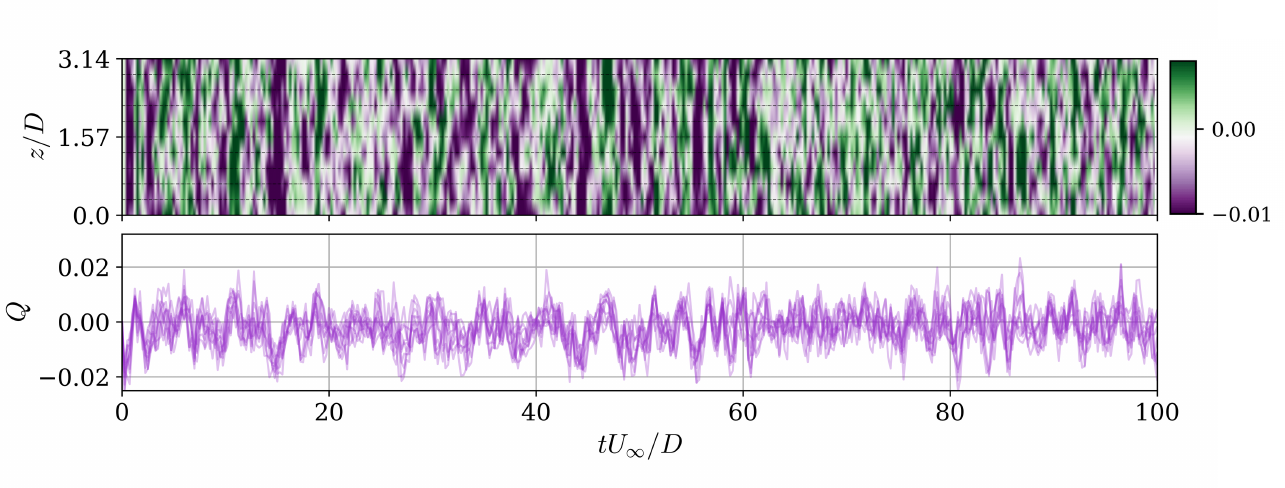}
        \begin{center}(b)\end{center}
    \end{minipage}
    \hfill
        \begin{minipage}[b]{0.49\textwidth}
        \centering
        \includegraphics[width=\textwidth,trim = 1.0cm 0.5cm 1.5cm 0.7cm, clip]{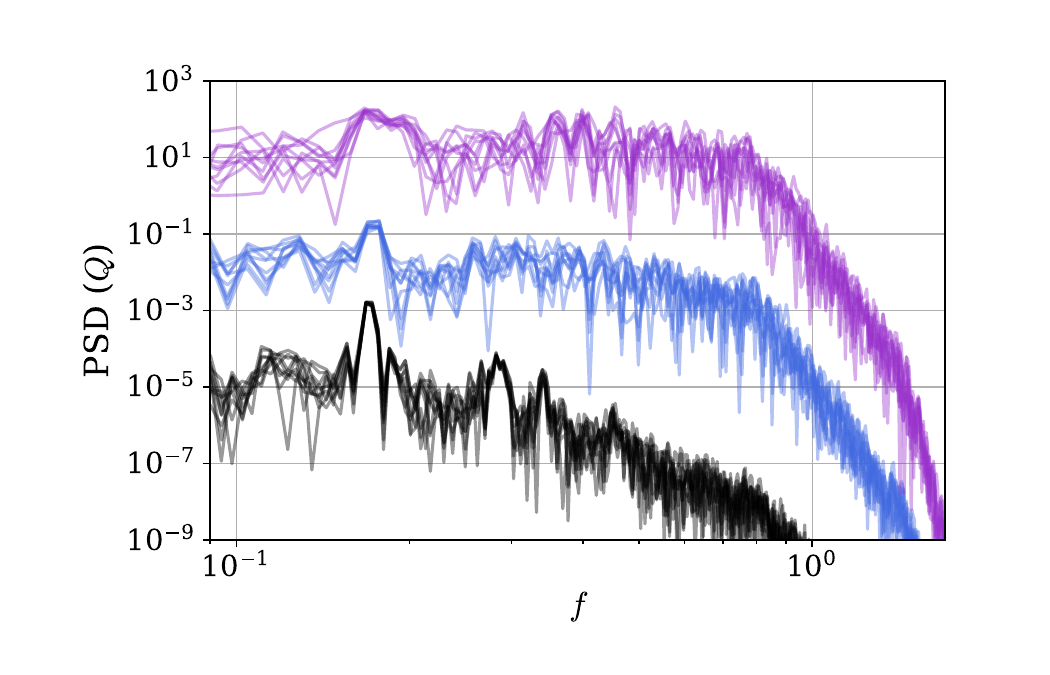}
        \begin{center}(c)\end{center}
    \end{minipage}
    \begin{minipage}[b]{0.49\textwidth}
        \centering
        \includegraphics[width=\textwidth,trim = 0.45cm 0 0.5in 0in, clip]{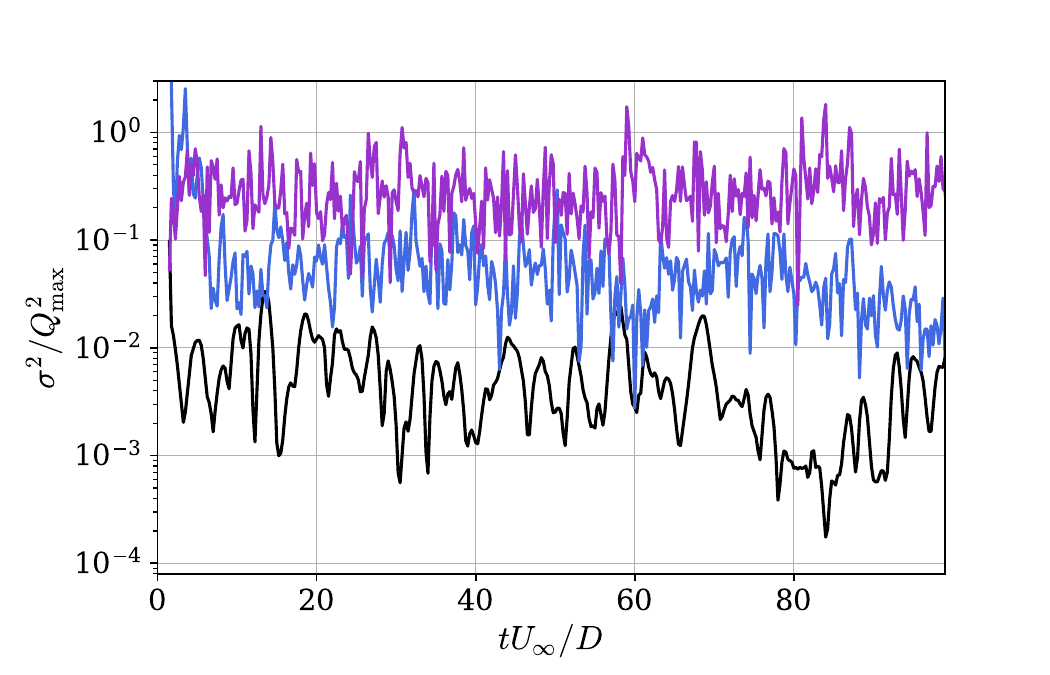}
        \begin{center}(d)\end{center}
    \end{minipage}
    \hfill
    \caption{\rev{Mass-flow rate per unit width $Q$ as a function of time for all jets, showing spatial distribution over $z/D$ (top) and Q for all jets overlapped (bottom) for (a) DRL-10-s1 and (b) DRL-10-s2. (c) Power-spectral density of all actuators $Q$ signals in time comparing (shifted by $10^2$ between cases, for visualization purposes). (d) Evolution in time of the variance of the mass-flow rate computed in $z$, $\sigma^2(t) = (1/n_{\rm{jets}}) \sum_{i=1}^{n_{\rm{jets}}} (Q_i(t)- \overline{Q}(t))^2$, and normalized by the squared peak $Q$ values from each case. Note that (c) and (d) show DRL-10-s1 (blue), DRL-10-s2 (purple) at $Re_D=3900$, and $Re_D=400$ reported in Ref.~\citep{suarez2023active} (black).}}
    \label{fig:pcolor}
\end{figure}

Figure \ref{fig:pcolor}(a) and \rev{\ref{fig:pcolor}(b) illustrate} the benefits offered by a system like MARL. In comparison to what is observed at lower Reynolds numbers~\cite{suarez2023active}, where it was clear that the control tries to synchronize all jets to act at the same frequency adapting to minimal instabilities, here we have a much richer control. As shown by the power-spectral density in Figure \ref{fig:pcolor}(c), $Q$ evolution undergoing DRL-based control exhibits more features besides the main frequency $f_c$. In contrast, strategy reported for $Re_D=400$ in Ref.~\cite{suarez2023active} only has a clear peak and a second harmonic compared to the rest of its spectrum in the figures. It can be observed a control resolution where multiple actuators collaborate. Among them, they form a distributed blowing/suction spanning a spanwise length of around $D$, which aligns with the wavelengths experienced in this classic fluid case. \rev{When comparing both DRL cases, it is interesting to observe how the bandwidth of the spectrum broadens after the second training stage, thus containing more frequencies around the $St$.}

\rev{As a general observation, building on the previous discussion about mass cost, it is worth noting that in the DRL strategies, the values are mainly within the range of $Q_{\rm s1/s2} \in [-0.01, 0.01]$, which is an order of magnitude smaller than in KC05, where the equivalent mass-flow per unit length is $Q_{\rm KC05} = 0.11$. Despite this, there are occasional peaks in $Q_{\rm s1}$ that saturate the signal, reaching $Q > 0.02$. This indicates that the DRL-10-s1 actuators are able to fully exploit the mass flow rate when needed. However, it also suggests that the system periodically loses track and effectively resets the flow state, repeatedly recovering a favorable reduction of $C_D$. This behavior is also reflected in the low-frequency fluctuations seen in the drag reduction signal already commented in Figure \ref{fig:cdandcl}. For DRL-10-s2, the evolution of $Q_{\rm s2}$ is slightly different: the sudden peaks are diminished, and the signal becomes more homogeneous in time, though more heterogeneous in the spanwise direction.}

Overall, it is important to highlight that despite the DRL agent being able to explore beyond this range during training, it decides to limit the maximum control values to be only $10\%$ of $Q_{\rm max}$ during exploitation. This suggests the algorithm is conservative, aiming to use the minimum mass-flow rate for maximum impact. In our experience, poor learning is often indicated by signal saturation, where the agent fails to adapt to the flow.

This sophisticated distribution is remarkable. If we focus on the standard deviations $\sigma$ shown in Figure \ref{fig:pcolor}(d), these are orders of magnitude higher than the those in the reference case at $Re_D=400$ reported in Ref.\cite{suarez2023active}. \rev{The agents at $Re_D=3900$ are capable of acting precisely on the dominant structures in the wake. Again, after the second stage \textit{i.e.} DRL-10-s2, the $\sigma$ the distribution over the span is richer than that in the  DRL-10-s2 case too.}

\begin{figure}[ht]
    \centering
    \begin{minipage}[b]{\textwidth}
        \centering
        \includegraphics[width=0.9\textwidth]{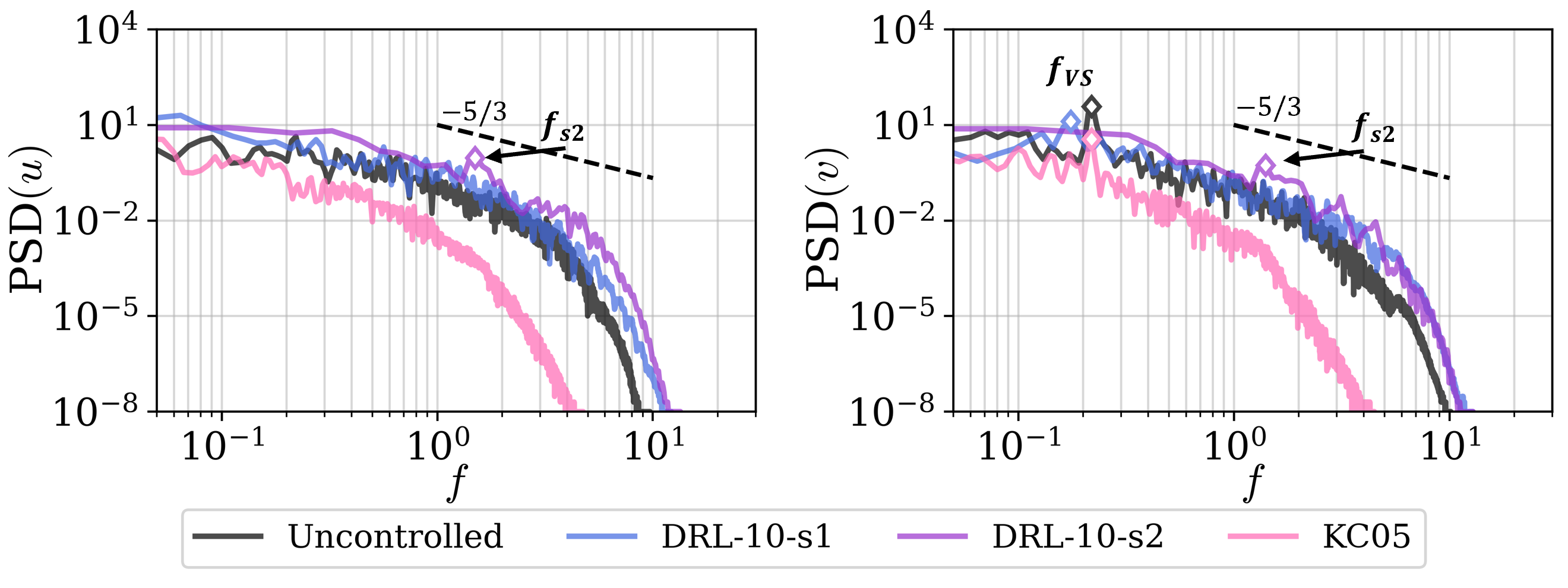}
        \begin{center}(a)\end{center}
        \label{fig:welch1}
    \end{minipage}
    \hfill
    \begin{minipage}[b]{\textwidth}
        \centering
        \includegraphics[width=0.9\textwidth]{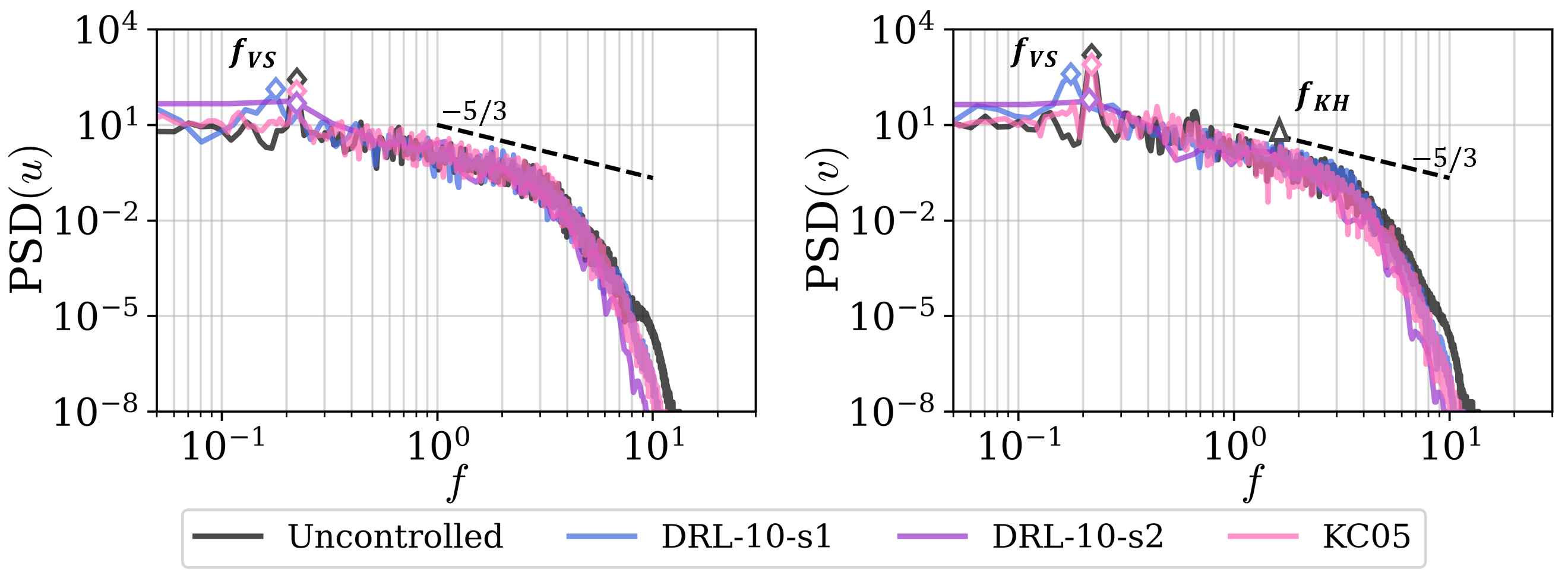}
        \begin{center}(b)\end{center}
        \label{fig:welch2}
    \end{minipage}
    \caption{Power spectral density of the (left) streamwise velocity $u$ and (right) cross-stream velocity $v$. The points considered are located at the following coordinates: (a) P1, $(x/D, y/D, z/D) = (6.81, 13.25, 1.26)$ and (b) P2, $(x/D, y/D, z/D) = (8.25, 13.25, 1.26)$. \rev{All studied cases are shown, with the main frequencies denoted by $f_{\rm VS}$ for vortex shedding and $f_{\rm KH}$ for Kelvin--Helmholtz instabilities. A dashed line representing the $E(k) \sim k^{-5/3}$ energy spectrum, characteristic of the inertial subrange in turbulent flows, is also included for reference.}}
    \label{fig:welch}
\end{figure}

Next, we study the spectra of temporal signals of the streamwise and cross-flow velocities for two locations in the wake of the cylinder, see Figure \ref{fig:welch}. Inspired by the study conducted by Lehmkuhl {\it et al.}~\cite{lehmkuhl2013low}, points P1 at $(x/D,y/D,z/D)=(6.81,13.25,1.26)$ and P2 at $(x/D,y/D,z/D)=(8.25,13.25,1.26)$ were chosen to assess the main frequencies and compare them with those reported in the literature.

In the uncontrolled case, the shedding frequency of vortices $f_{\rm vs}=0.22$ is clearly captured at both locations. However, at the P1 location, it seems to be too close to the cylinder surface to effectively capture the bubble instability, especially when considering the streamwise velocity. Additionally, at point P2, we can discern the emergence of a higher frequency at $f_{\rm KH}=1.55$, which according to literature, could be associated with the Kelvin--Helmholtz instability in the separating shear layer. The observed value is slightly higher than the estimate provided by Prasad and Williamson \cite{PRASAD_WILLIAMSON_1997}, which is given as $f_{\rm KH} = 0.0235Re_D^{0.67}f_{\rm vs} = 1.31$.

If we compare with the results of the controlled cases, we can observe the influence of the actuators in both DRL-based and KC05. In DRL-10-s1, there is a shift to a lower shedding frequency and lower intensity, $f_{\rm vs}^{s1}=0.177$. \rev{On the other hand, DRL-10-s2 turns to decrease the intensity of those fluctuations on those locations--indicating that the recirculation bubble is a bit larger than DRL-10-s1.} Meanwhile, KC05 maintains the same shedding frequency, $f_{\rm vs}^{KC05}=f_{\rm vs}=0.22$, but with a slight decrease in intensity. This is consistent with the fact that the KC05 control does not involve any temporal dependency, thus not affecting the dominant frequency in the wake. At point P1 (Figure \ref{fig:welch}), it is interesting how the intensity of the energy cascade for KC05 is much lower compared with to the uncontrolled case, while \rev{DRL-10-s1/s2} exhibit a slight increase. It can be inferred that the agent is enriching the finer scales near the cylinder, whereas the strong actuation of KC05 causes such structures to fade. \rev{Note that $St$ values shown in Table~\ref{tab:results} are computed using spectra from probes located further away, as well as from the lift signal.}

\begin{figure}[ht]
    \centering
    \begin{minipage}[b]{0.99\textwidth}
        \centering
        \includegraphics[width=\textwidth]{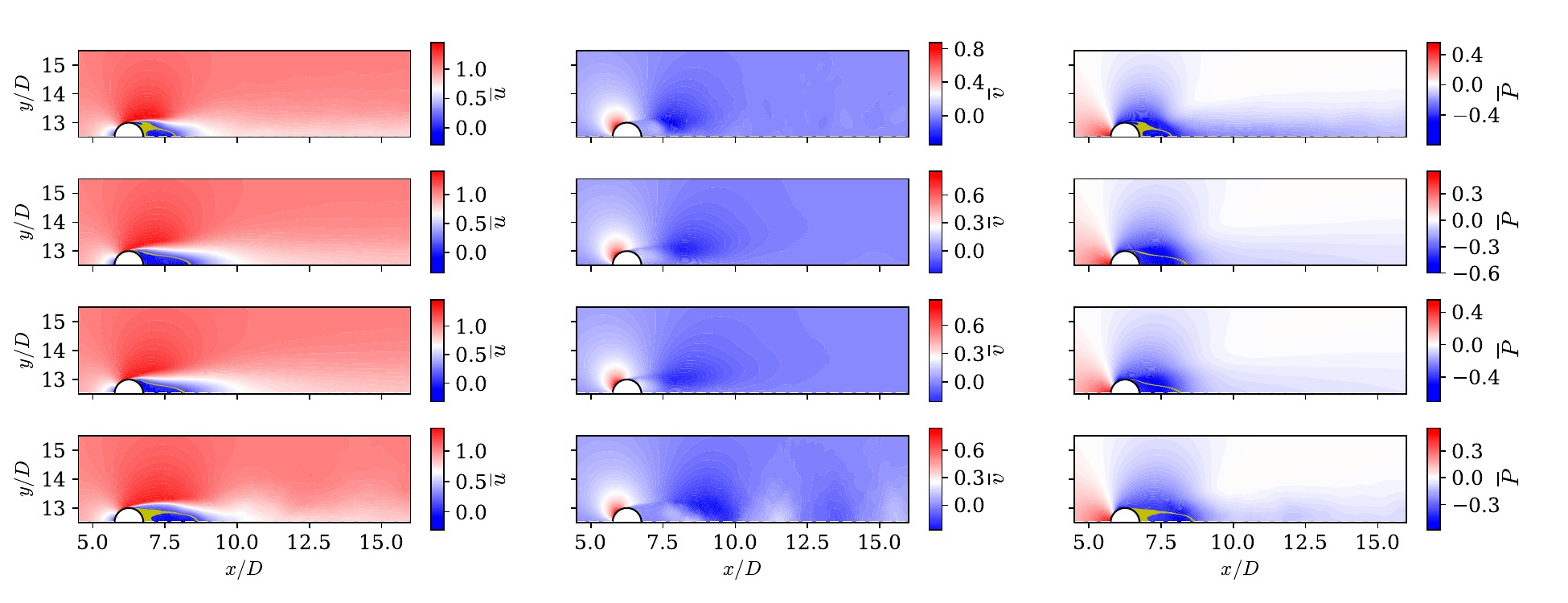}
        \begin{center}(a)\end{center}
    \end{minipage}
    \hfill
    \begin{minipage}[b]{0.99\textwidth}
        \centering
        \includegraphics[width=\textwidth]{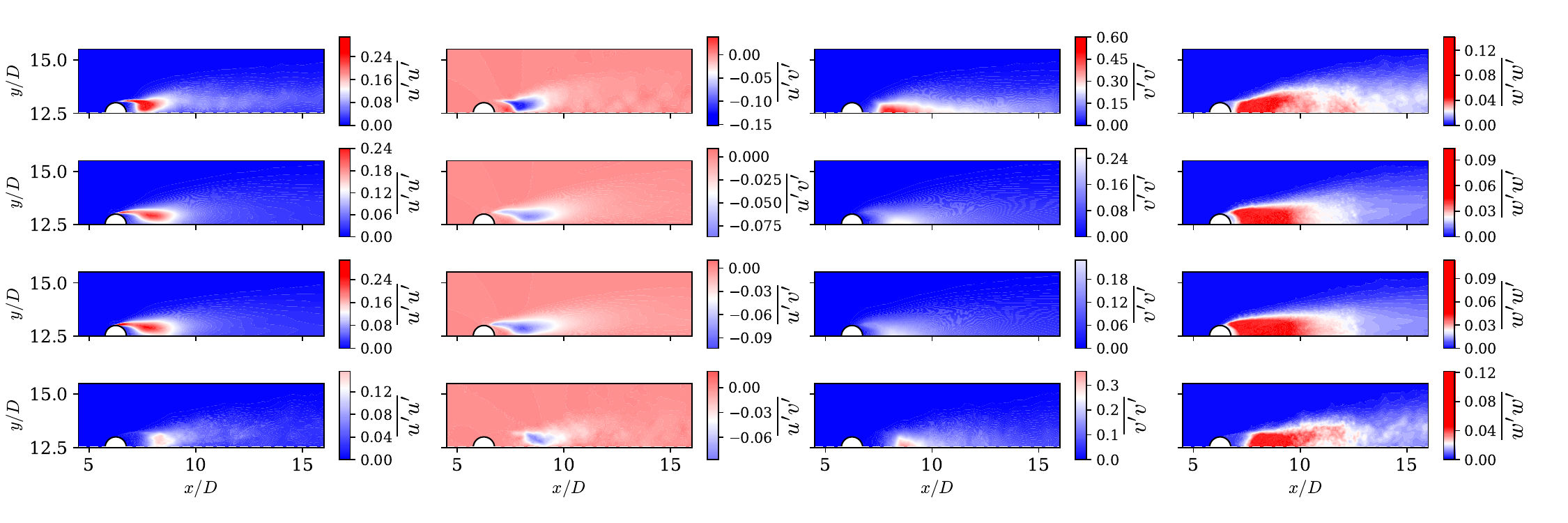}
        \begin{center}(b)\end{center}
    \end{minipage}
        \begin{minipage}[b]{\textwidth}
        \centering
        \includegraphics[width=0.95\textwidth]{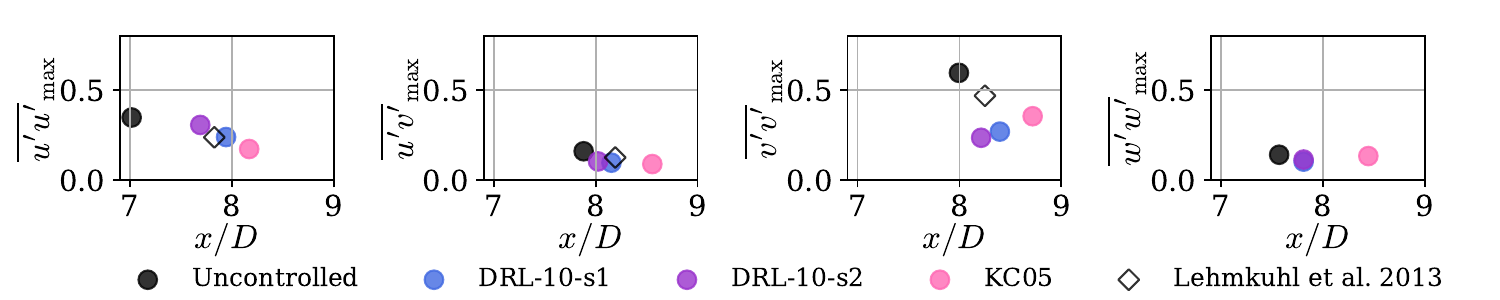}
        \begin{center}(c)\end{center}
    \end{minipage}
    \caption{Time- and spanwise-averaged flow-field statistics comparison\rev{--averaged all over 150 time units after transients.} (a) Mean velocities, $\overline{u}$ and $\overline{v}$, and mean pressure $\overline{p}$ fields for \rev{from top to bottom: uncontrolled, DRL-10-s1, DRL-10-s2 and KC05.} Yellow regions denote wake-stagnation points where $|\overline{u}| <0.03$, which is used to compute $L_r/D$ in Table \ref{tab:results}. (b) Mean Reynolds stresses $\overline{u'u'}, \overline{u'v'}, \overline{v'v'}$ and $\overline{w'w'}$ for \rev{from top to bottom: uncontrolled, DRL-10-s1, DRL-10-s2 and KC05.} (c) Maximum Reynolds-stress values and their corresponding $x/D$ locations across all investigated cases, compared with values from the literature~\cite{lehmkuhl2013low}.}
    \label{fig:avg_maps}
\end{figure}

The influence of the studied AFC is not limited to frequencies but also extends to the mean fields in the wake. First, Figure~\ref{fig:avg_maps}a shows how the recirculation bubble is enlarged as a result of both DRL controls. In DRL-10-s1 and DRL-10-s2, we can observe very similar results: a shorter recirculation bubble than in the KC05 case, also exhibiting fewer regions of mean streamwise velocity where $|\overline{u}| <0.03$ (highlighted as yellow-colored regions). The KC05 case resembles the uncontrolled configuration, with larger regions of zero velocity than the DRL cases. 

Second, in Figure~\ref{fig:avg_maps}(b) and Figure~\ref{fig:avg_maps}(c) we analyze the Reynolds stresses. We observe a similar trend in all controlled cases, where the maximum stresses are reduced by $\approx 50\%$ and their locations shifted upstream. \rev{Note that the KC05 case prominently exhibits the shift of the locations of those peaks by a length of around $D$. For DRL-based control, the change is less pronounced but still effective, as the absolute values of the stresses decrease. Specifically, the $\overline{v'v'}$ component reduces by more than 60\% compared to the baseline values.} This behavior is consistent with the drag-reduction mechanisms observed at lower Reynolds number~\cite{suarez2023active}.

\begin{figure}[ht]
    \centering
    \includegraphics[width=\textwidth, trim = 0cm 0 0cm 0, clip]{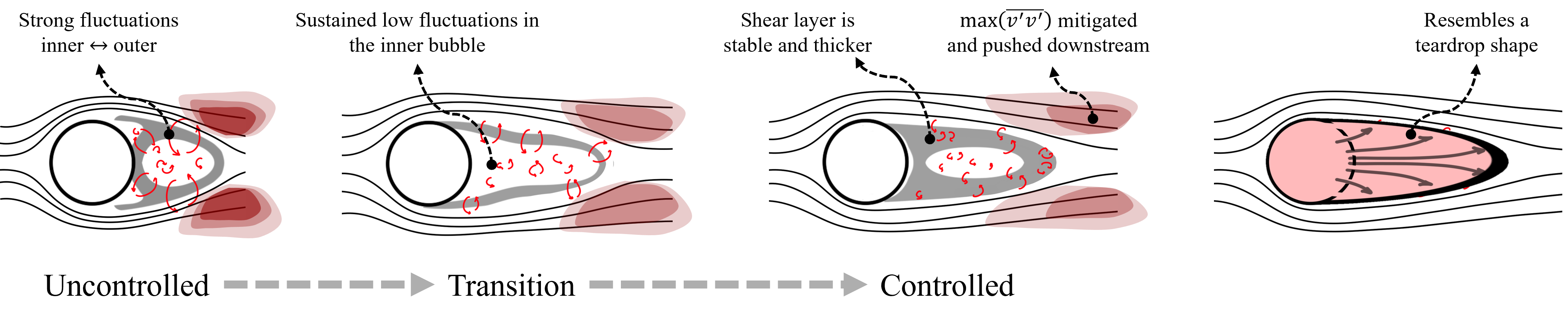}
    \caption{\rev{Schematic representation of wake topology evolution, illustrating the transition from uncontrolled to controlled flow states. The diagram highlights the key drag reduction mechanisms and their influence on wake dynamics and flow organization.}}
    \label{fig:scheme_dragredu}
\end{figure}

\rev{This is consistent with previous findings for similar configurations. Specifically, the DRL agents work to expand the recirculation area and suppress mixing between the inner and outer flow within the recirculation bubble. This results in a more stable and extended outer shear layer enveloping the bubble. Doing so, the actuators effectively reshape the flow into a more aerodynamically efficient profile, resembling a teardrop or airfoil, which is known to minimize drag. Figure \ref{fig:scheme_dragredu} provides a simplified depiction of how these drag reduction mechanisms evolve and influence the wake topology.}

\begin{figure}[ht]
\begin{minipage}[b]{0.49\textwidth}
        \centering
        \includegraphics[width=\textwidth]{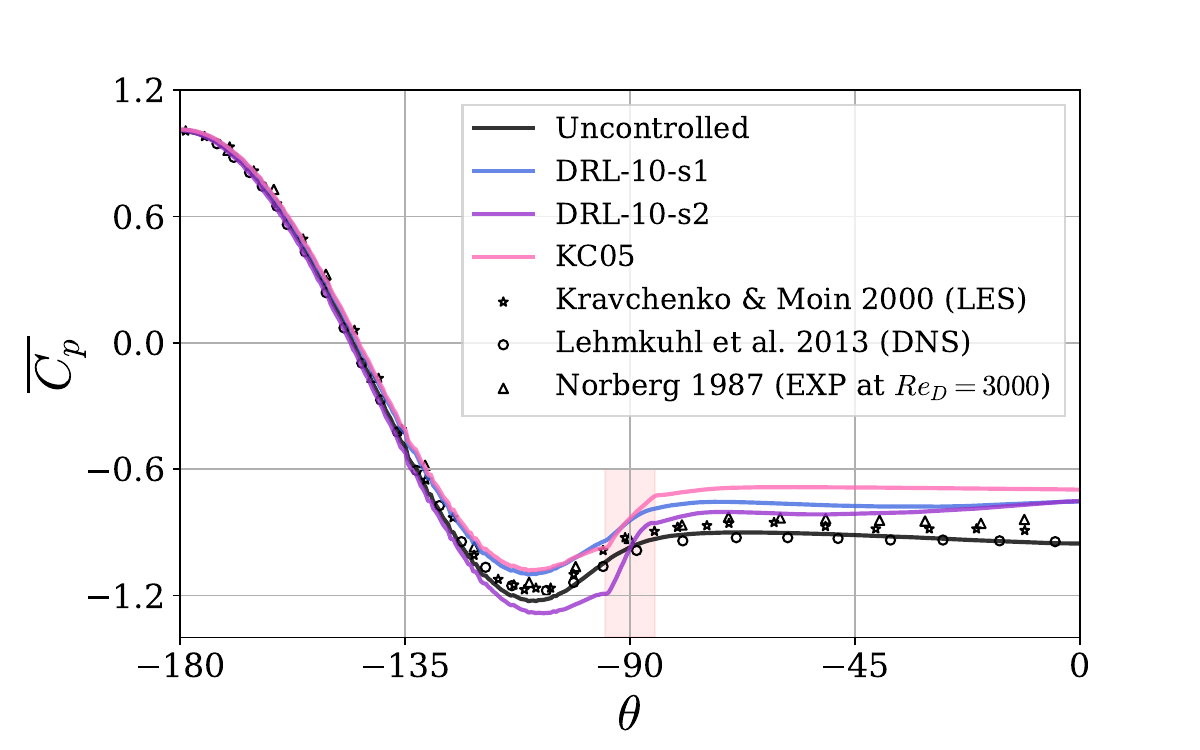}
        \begin{center}(a)\end{center}
    \end{minipage}
    \begin{minipage}[b]{0.49\textwidth}
        \centering
        \includegraphics[width=\textwidth, trim = 0 0 0 0.2cm, clip]{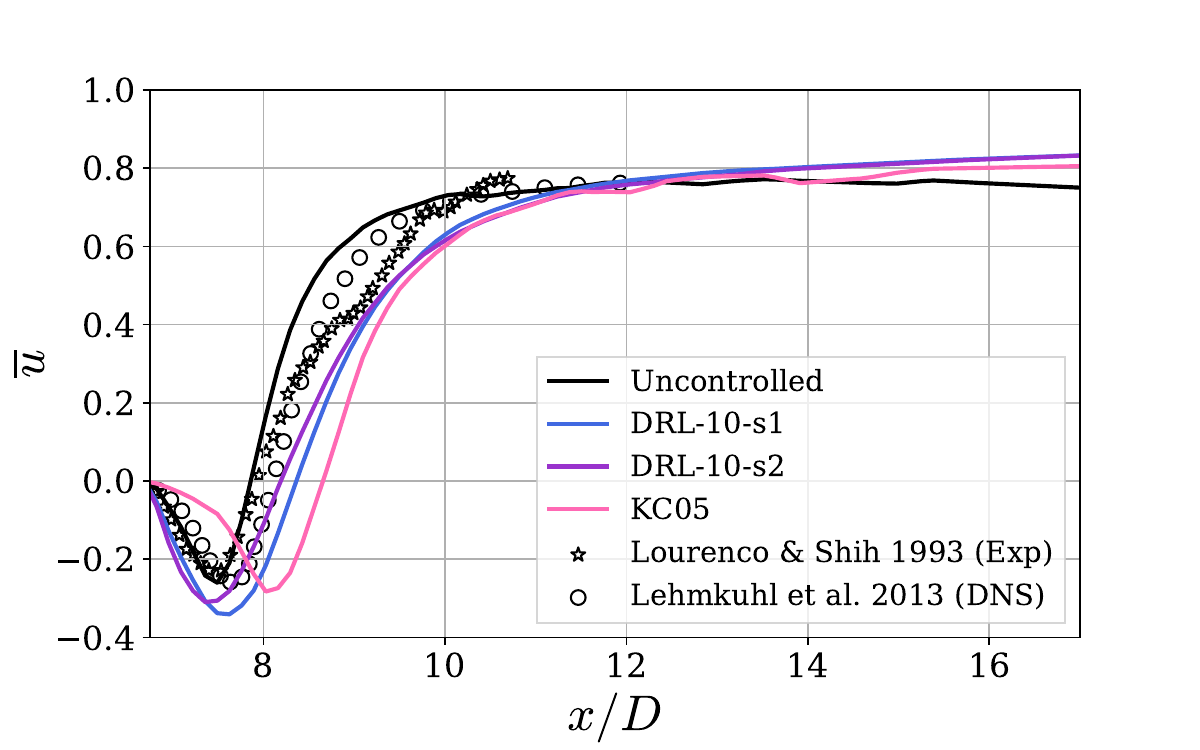}
        \begin{center}(b)\end{center}
    \end{minipage}
    \caption{(a) Pressure-coefficient distribution around the cylinder, where $\theta=0$ is the back of the cylinder. \rev{The red shaded area corresponds to the location of the jets, that spans $\omega=10^\circ$.} (b) Averaged streamwise velocity along the centerline at $y/D=L_y/2$. Note that several curves from the literature are included to validate the present uncontrolled case in (a) and (b).}
    \label{fig:press_distr}
\end{figure}

The pressure distribution on the surface of the cylinder is shown to be consistent with the literature, except for a slight deviation in the minimum peak at $\theta \approx -112.5^\circ$, as seen in the Figure \ref{fig:press_distr}(a). \rev{However, when considering the influence of all controlled cases, around this pressure valley, we observe that both DRL-10-s1 and KC05 follow the same trend, recovering some pressure. On the other hand, unexpectedly, DRL-10-s2 deviates from these strategies, further decreasing this valley value.} Nevertheless, we observe that the most significant change occurs in the distribution, specifically within the separation bubble delimited by $\pm \theta_s = 89^\circ$.

\rev{In the Figure \ref{fig:press_distr}(b), we observe a V-shaped mean streamwise velocity along the centerline in the wake when uncontrolled, which shows good agreement with the literature. The controlled cases, however, exhibit a more U-shaped velocity profile. The reattachment, with the downstream velocity value--which is more relevant--occurs later for all controlled cases, indicating an increase in the recirculation bubble, as discussed earlier in Table~\ref{tab:results}. Both DRL-10-s1 and DRL-10-s2 reach higher negative velocity values, while KC05 is better at maintaining regions closer to $\overline{u}=0$ near the cylinder. The minimum velocity values are also noticeably shifted by the latter, around $D/2$.}

\begin{figure}[ht]
        \centering
        \includegraphics[width=\textwidth, trim= 0.5cm 0 0.5cm 1cm, clip]{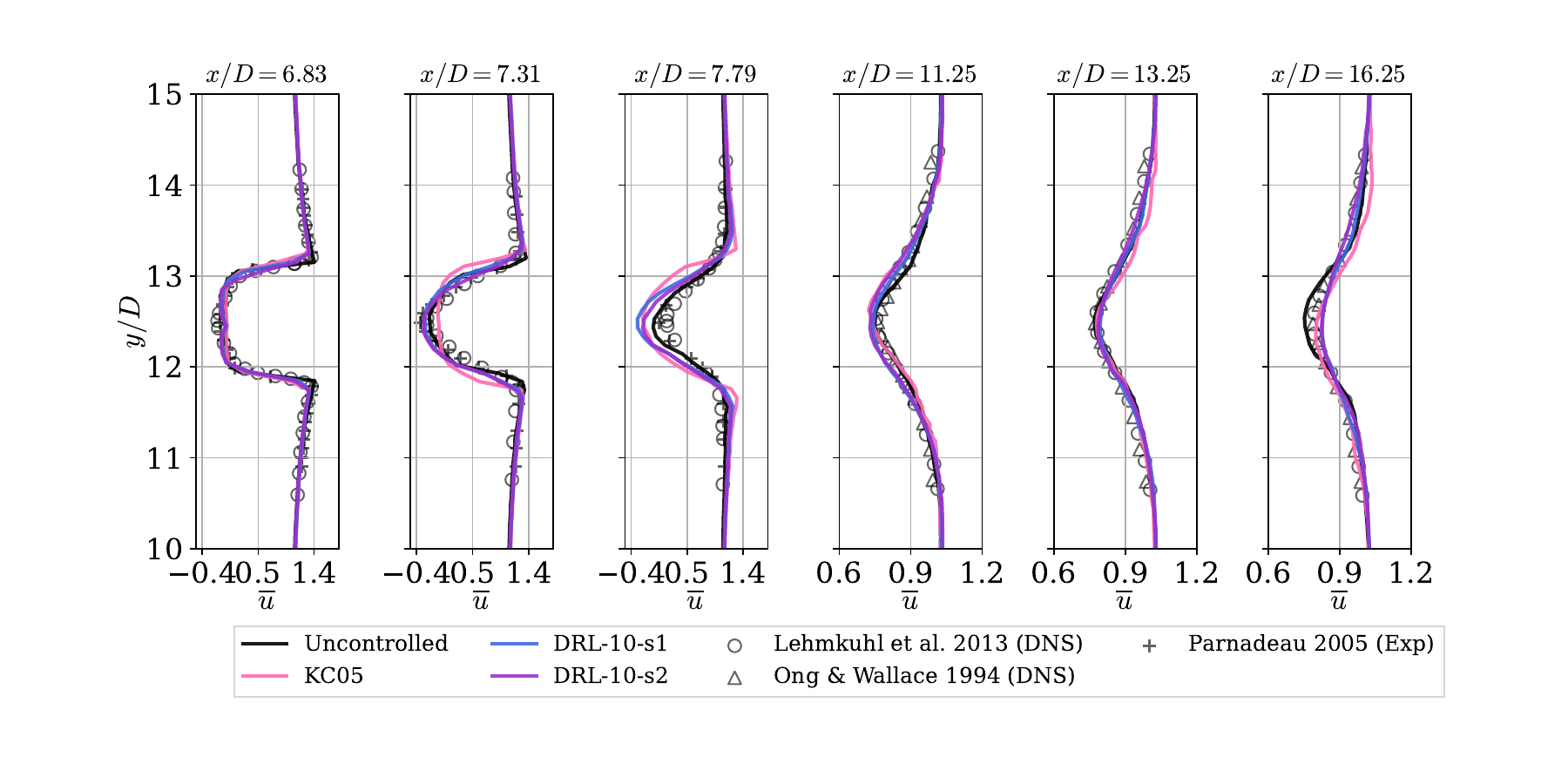}
    \caption{Mean streamwise velocity $\overline{u}$ profiles at different locations along the wake. Note that reference trends from the literature are included to validate the present uncontrolled case in all the figures.}
    \label{fig:wakes}
\end{figure}

To conclude the analysis, the different velocity profiles along the wake show very good agreement with the literature, see Figure \ref{fig:wakes}. Given that significant scattering with the references typically happens near the cylinder, the observations at the first three locations ($x/D=6.83$, 7.31, and 7.79) reaffirm our earlier observations: \rev{The recirculation zone is wider in the controlled cases, with KC05 showing a slightly more pronounced effect compared to the DRL cases. Additionally, $\overline{u}$ exhibits a flatter profile near the centerline. For instance, at $x/D=7.79$, it is noteworthy how the recirculation zone is maintained in the controlled cases, while in the uncontrolled case, $\overline{u}$ reattaches to the downstream velocity further into the wake. This indicates that the velocity deficit remains higher, and the momentum is more sustained in the controlled cases. As expected, farther from the cylinder, the wake recovers its shape similarly across all controlled cases, with no significant differences observed in the redistribution of momentum from the central velocity deficit to the upper or lower regions.}

\section{Conclusions}\label{sec_conclusion}
This study couples a multi-agent reinforcement learning (MARL) framework with a numerical solver to explore efficient drag-reduction strategies by controlling multiple jets positioned along the span of a three-dimensional cylinder. The investigation is conducted at $Re_D=3900$, representing the fully turbulent wake, and compared with a classical controlled case based on the \textit{out-of-phase} strategy reported by Kim~$\&$~Choi~\cite{choi_distributed_2005}, referred to as KC05. \rev{The framework benefits from an innovative multi-stage training workflow for AFC applications, where training is optimized dynamically. In this process, the policy $\pi$ is evaluated at two stages: after the first phase, DRL-10-s1, and after the second phase, DRL-10-s2.} 

The DRL-based control policies outperform the ratio of mass cost per drag reduction rate used in the KC05 strategy by two orders of magnitude. Both DRL-10-s1 and DRL-10-s2 exploit the emergence of spanwise instabilities to discover successful \rev{drag-reduction strategies with $\overline{\Delta C_{D_{s1}}}=-9.4\%$ and $\overline{\Delta C_{D_{s2}}}=-8.3\%$ respectively.} This is achieved by leveraging the underlying physics within pseudo environments and optimizing the global problem involving multiple interactions concurrently. \rev{Notably, the cooperative closed-loop strategy developed by the agents introduces a novel approach, utilizing a range of mass-flow-rate frequencies that would be difficult to achieve using classical control methods.} These findings underscore the DRL approach’s ability to identify more refined flow-control strategies than those previously achieved through classical methods, covering a broad range of frequencies and effectively managing various flow-structure wavelengths in the wake.
\rev{Regarding the drag-reduction mechanisms, the actuators discover a novel way to interact with the flow around the three-dimensional cylinder by enlarging the recirculation bubble, influencing the fluctuations by shifting them downstream and attenuating their intensity. At the same time, they attenuate the mixing between the inner and outer region of the bubble. As a result, the wake behind the cylinder takes on a shape resembling a teardrop or airfoil-like topology, which is known to be more efficient for drag reduction (as illustrated in Figure \ref{fig:scheme_dragredu}).}

Another advantage of MARL is its ability to deploy trained agents across various cylinder lengths and numbers of actuators, while ensuring consistency in the spanwise width of the jets and the corresponding pressure measurement locations as observation states. \rev{It is worth noting that the training focuses on symmetries and invariant structures, a task not feasible with SARL due to its limitations in scaling to different geometries}. MARL enables computational cost-effective training sessions in smaller and simplified computational domains, thereby accelerating the process required for flow control in high-fidelity simulations. \rev{This, combined with the results from implementing the second stage to address weaknesses already observed in the DRL-10-s1 policy, suggests that there are many ways to enrich the exploration process and enhance the capabilities of the same training session without having to start from scratch, as seen with the DRL-10-s2.}

\rev{The current DRL-based AFC strategies demonstrate impressive efficiency in reducing drag, offering both immediate energy savings through on-the-fly adjustments and long-term actuation benefits. A particularly novel and important contribution of these strategies is their non-invasive nature, which allows significant modifications to the wake topology while maintaining minimal disruption to existing systems. From a physics perspective, this non-invasiveness is especially interesting, as it enables substantial changes in the flow dynamics. Furthermore, this study is the first to perform training on a fully turbulent 3D cylinder flow at \( Re_D = 3900 \), within the framework of MARL with two-stage exploration. This achievement provides a valuable reference for the DRL community and may encourage future applications in distributed-input, distributed-output systems.

While the results are promising, the practical implementation of these strategies in real-world applications will require careful consideration of computational resource demands and the ability for real-time adjustments. Further exploration is needed to assess the feasibility of DRL-based strategies for industrial use.
}

\newpage

\appendix
\renewcommand{\thesection}{Appendix \Roman{section}}  
\section{: Grid convergence study} \label{App1}  

\begin{figure}[ht]
        \centering
        \includegraphics[width=0.9\textwidth]{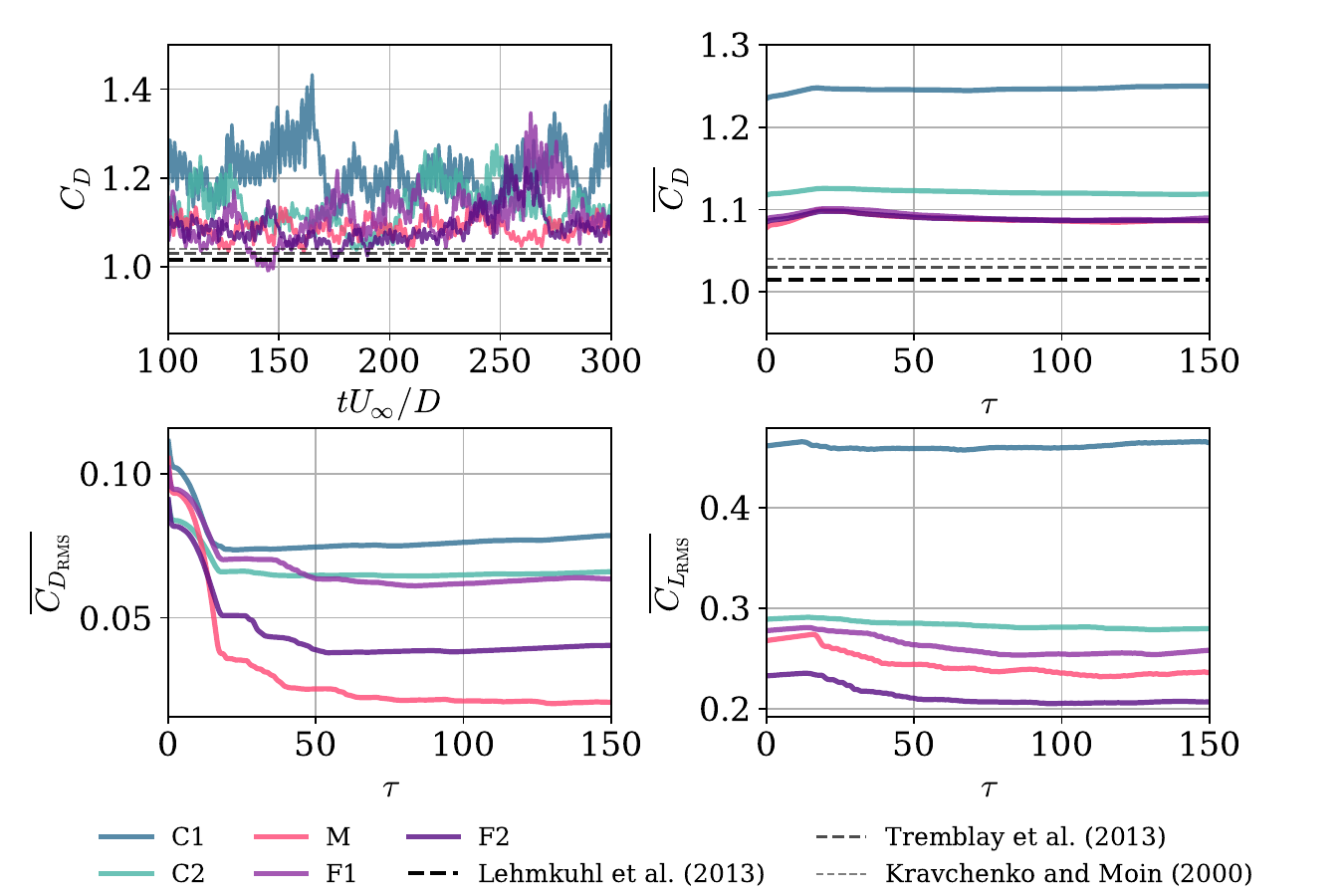}  \caption{\rev{Convergence analysis for $C_D$ and $C_L$ signals for the 5 different proposed meshes. (top left) Temporal evolution $C_D=f(t)$. (top right) Accumulative average, $\overline{C_D}=f(\tau)$ and (bottom left and bottom right) RMS for $C_D$ and $C_L$ as $f(\tau)$ aswell, respectively; starting from the last timestep and advancing backwards where $\tau=(t_{\rm end}-t)U_{\infty}/D$ and $\overline{\phi}(\tau)=(1/\tau)\int^{\tau}_0 \phi(t) dt$.}}
    \label{fig:convergence}
\end{figure}

\begin{figure}[ht]
       \centering
       \includegraphics[width=0.6\textwidth, trim = 3cm 6cm 8cm 6cm, clip]{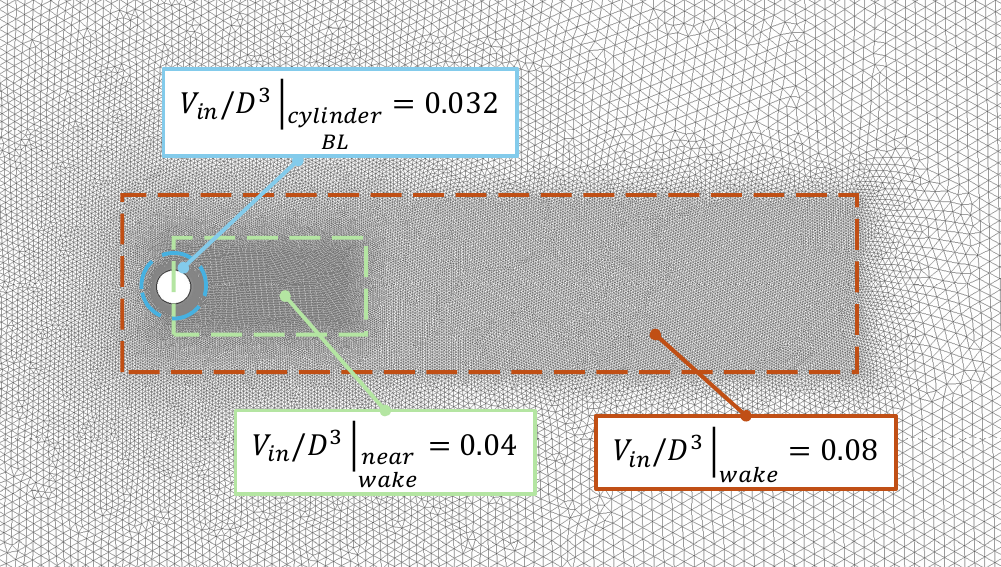}
       \caption{\rev{Mesh used in M case with details regarding the refinement boxes. Where $V_{\rm in}/D^3$ is set in Gmsh software.}}
       \label{fig:mesh}
\end{figure}

\rev{In order to ensure the reliability and accuracy of the results, a grid independence study was conducted for three-dimensional cylinders at $Re_D=3900$, with the drag coefficient as the primary integral quantity of interest. The study reveals that the drag coefficient converges as the mesh is refined, stabilizing with finer grids, as shown in Table \ref{tab:grid_indepe}. In this table, meshes M, F1, and F2 converge around $\overline{C_D} = 1.08$. We note that our results are slightly higher values than those reported in the literature, even with much finer meshes. Although the literature also reports modes that alternate between $C_D = 1.00$ and $1.08$ and reinforces the validity of our findings.

Figure \ref{fig:convergence} illustrates how all averaged quantities reach a plateau when $\tau > 100$. Additionally, other key statistics, such as mean--flow patterns and pressure distributions, were validated for the used mesh and are shown in Figures \ref{fig:press_distr} and \ref{fig:wakes}, as well as in Table \ref{tab:validate}. Note that is important to indentify the minimum number of grid points that would provide adequate accuracy. This study strikes a balance between the desired simulation quality and the high computational demand typical of DRL training, ensuring reliable results for large-scale simulations while maintaining the accuracy of the drag coefficient and other flow characteristics.

As a final comment on the mesh depicted in Figure \ref{fig:mesh}, it is generated by a parametric GMSH script with three distinct refinement regions: the boundary layer around the cylinder surface, the near wake, and the larger wake.}

\begin{table}[ht]
\caption{\rev{Summary of grid independence study using five different mesh for $Re_D=3900$: coarse (C), medium (M), and fine (F). The medium mesh (M) was used in the present paper. All quantities are averaged over the last 150 time units of the simulation. We refer to table \ref{tab:validate} for some references to compare them.}}
\label{tab:grid_indepe}
\begin{tabular}{ccccccc}
\toprule
Case & Grid points & $\rm{min}(\Delta r)$ & $\Delta z$ & $\overline{C_D}$ & $\overline{C_{D_{\rm RMS}}}$ & $\overline{C_{L_{\rm RMS}}}$ \\
\midrule
    C1 & $7.6 \times 10^6$ & 0.0049 & 0.0314 & 1.240 & 0.078 & 0.464 \\
    C2  & $9.2 \times 10^6$ & 0.0045 & 0.0260 & 1.119 & 0.066 & 0.279 \\
    M & $9.6 \times 10^6$ & 0.0036 & 0.0260 & 1.081 & 0.021 & 0.236 \\
    F1 & $13.5 \times 10^6$ & 0.0031 & 0.0260 & 1.089 & 0.063 & 0.258 \\
    F2 & $17.8 \times 10^6$ & 0.0026 & 0.0209 & 1.083 & 0.041 & 0.201 \\
\botrule
\end{tabular}
\end{table}

\pagebreak
\newpage

\section*{Declarations}

\subsection*{Funding}
This study was enabled by resources provided by the National Academic Infrastructure for Supercomputing in Sweden (NAISS) at PDC, KTH Royal Institute of Technology. R.V. acknowledges financial support from ERC grant no.2021-CoG-101043998, DEEPCONTROL. Views and opinions expressed are however those of the author(s) only and do not necessarily reflect those of the European Union or the European Research Council. Neither the European Union nor the granting authority can be held responsible for them.

\subsection*{Conflict of interest}
The authors have no conflicts to disclose.

\subsection*{Ethical approval}
Not applicable.

\subsection*{Informed consent}
Not applicable.

\subsection*{Authors' contributions}

\textbf{Suárez, P.:}  Methodology, software, validation, investigation, writing - original draft and visualization. \textbf{Alcántara-Ávila, F., Rabault, J., Miró, A. \& Font, B. :} Methodology, software, and writing - review \& editing. \textbf{Lehmkuhl, O. :} Funding acquisition, supervision, and writing - review \& editing. \textbf{Vinuesa, R.:} Conceptualization, project definition, methodology, resources, writing - original draft, supervision, project administration and funding acquisition.

\subsection*{Data availability statement}
The data presented in this study are available upon request from the corresponding~author.

\bibliography{biblio_pol}

\end{document}